\documentclass[twocolumn,english]{article}
\usepackage{charter}
\usepackage[T1]{fontenc}
\usepackage[latin9]{inputenc}
\usepackage{geometry}
\usepackage{babel}
\usepackage{float}
\usepackage{multirow}
\usepackage{amsmath}
\usepackage{graphicx}
\usepackage[authoryear]{natbib}
\usepackage{subscript}
\usepackage[unicode=true,
 bookmarks=false,
 breaklinks=true,pdfborder={0 0 0},pdfborderstyle={},backref=false,colorlinks=false]
 {hyperref}

\makeatletter

\providecommand{\tabularnewline}{\\}

\newcommand{\lyxaddress}[1]{
	\par {\raggedright #1
	\vspace{1.4em}
	\noindent\par}
}

\usepackage{babel}
\usepackage{cuted}
\usepackage{tikz}

\makeatother

\begin{document}
\onecolumn
\thispagestyle{empty} 
\begin{center}
\textsc{\LARGE{}Scaling, mirror symmetries and consonances among the
distances of the planets of the solar system}{\LARGE\par}
\par\end{center}

\vspace{0.5cm}

\begin{center}
Michael J. Bank and Nicola Scafetta
\par\end{center}

\begin{center}
\begin{figure}[H]
\centering{}\includegraphics[width=1.05\textwidth]{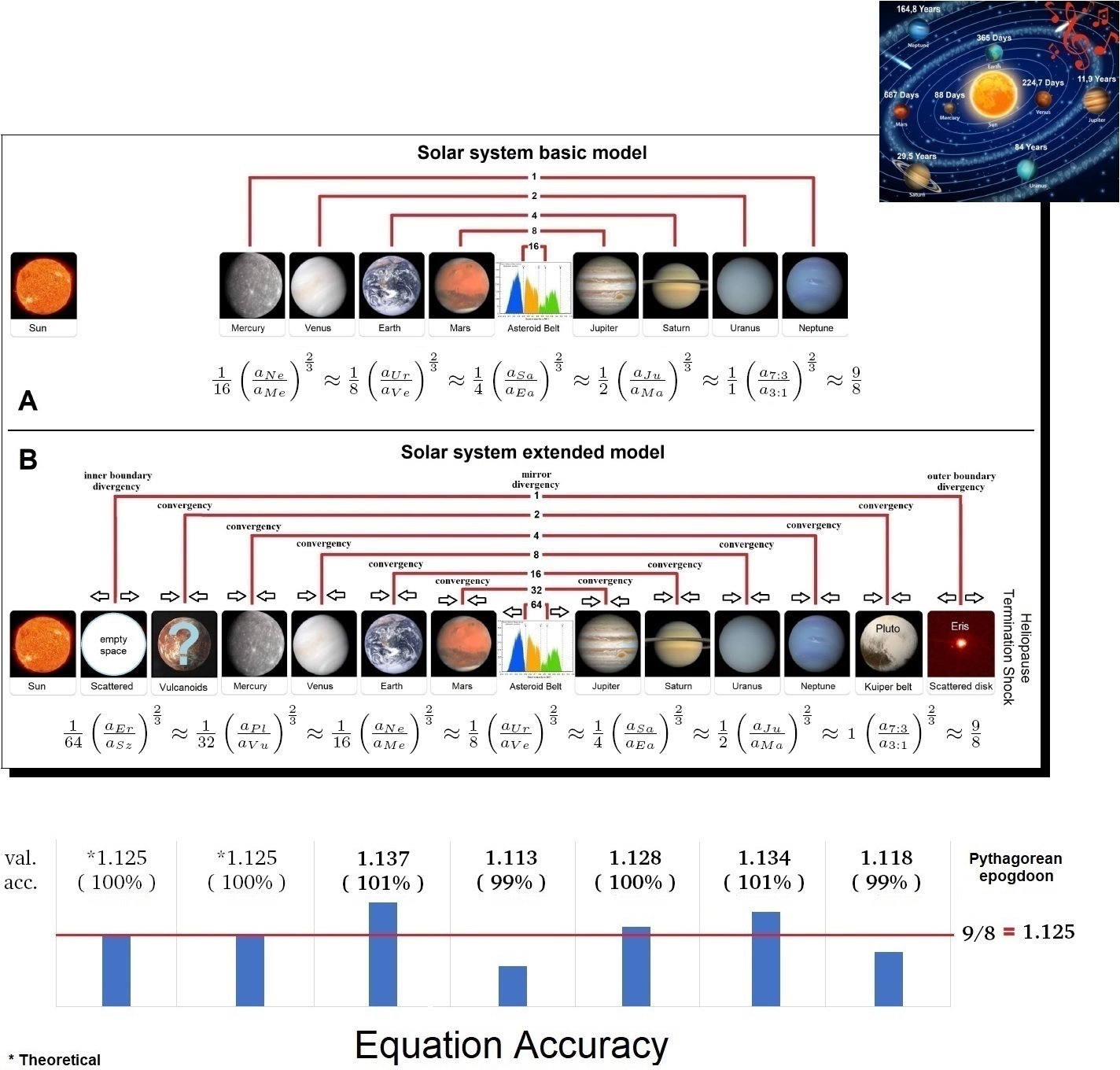}
\end{figure}
\par\end{center}

\noindent \textbf{Citation:} Bank M. J., Scafetta N.: 2022. Scaling,
Mirror Symmetries and Musical Consonances Among the Distances of the
Planets of the Solar System. \emph{Frontiers in Astronomy and Space
Sciences} 8:758184. \href{https://doi.org/10.3389/fspas.2021.758184}{https://doi.org/10.3389/fspas.2021.758184} 

\newpage{}
\title{Scaling, mirror symmetries and consonances among the distances of
the planets of the solar system}
\author{Michael J. Bank\textsuperscript{1\dag{}} and Nicola Scafetta\textsuperscript{2\dag {*}}}
\maketitle

\lyxaddress{\textsuperscript{1}Danbury Music Centre, Danbury, CT, 06810 USA..}

\lyxaddress{\textsuperscript{2}Department of Earth Sciences, Environment and
Georesources, University of Naples Federico II, Complesso Universitario
di Monte S. Angelo, via Cinthia, 21, 80126 Naples, Italy.}

\lyxaddress{\textsuperscript{\dag{}}These authors share first authorship.}

\lyxaddress{\textsuperscript{{*}}Corresponding author: nicola.scafetta@unina.it }
\begin{abstract}
Orbital systems are often self-organized and/or characterized by harmonic
relations. Inspired by music theory, we rewrite the \citet[QJRAS, 24, 10--13,][]{Geddes}
equations for mirror symmetries among the distances of the planets
of the solar system in an elegant and compact form by using the 2/3rd
power of the ratios of the semi-major axis lengths of two neighboring
planets (eight pairs, including the belt of the asteroids). This metric
suggests that the solar system could be characterized by a scaling
and mirror-like structure relative to the asteroid belt that relates
together the terrestrial and Jovian planets. These relations are based
on a 9/8 ratio multiplied by powers of 2, which correspond musically
to the interval of the Pythagorean \emph{epogdoon} (a Major Second)
and its addition with one or more octaves. Extensions of the same
model are discussed and found compatible also with the still hypothetical
vulcanoid asteroids versus the transneptunian objects. The found relation
also suggests that the planetary self-organization of our system could
be generated by the 3:1 and 7:3 resonances of Jupiter, which are already
known to have shaped the asteroid belt. The proposed model predicts
the main Kirkwood asteroid gaps and the ratio among the planetary
orbital parameters with a 99\% accuracy, which is three times better
than an alternative, recently proposed harmonic-resonance model for
the solar system. Furthermore, the ratios of neighboring planetary
pairs correspond to four musical ``consonances'' having frequency
ratios of 5/4 (Major Third), 4/3 (Perfect Fourth), 3/2 (Perfect Fifth)
and 8/5 (Minor Sixth); the probability of obtaining this result randomly
has a $p<0.001$. Musical consonances are ``pleasing'' tones that
harmoniously interrelate when sounded together, which suggests that
the orbits of the planets of our solar system could form some kind
of gravitationally optimized and coordinated structure. Physical modeling
indicates that energy non-conserving perturbations could drive a planetary
system into a self-organized periodic state with characteristics vaguely
similar of those found in our solar system. However, our specific
finding suggests that the planetary organization of our solar system
could be rather peculiar and based on more complex and unknown dynamical
structures. \\
\end{abstract}
\\
\textbf{Note:} The present copy slightly updates the published paper
with the addition of Table 7 and relative comment on pages 15 and
16, with the artwork of the solar system depicted in Figure 8 (page
19) and the Appendix on page 20.\\
\\
\setcounter{page}{1}
\twocolumn

\section{Introduction}

Since ancient times the stability of the solar system, its regularities,
and the movements of its planets have attracted the attention of astronomers
and philosophers because their orbital revolutions appeared to be
related by simple proportions \citep{Haar,Stephenson1974,Godwin}.
This observation yielded the understanding that solar and lunar systems
could be characterized by harmonic resonances, which are usually the
result of self-organizing gravitational or tidal processes yielding
to long-term stable planet and moon orbits \citep{Aschwanden,Moons}.
In fact, planetary circular orbits are dynamically unstable, unless
their mutual orbital periods fall into harmonic whole number ratios
called ``orbital commensurabilities'' \citep{McFadden,Peale,Pakter}.

For example, within our solar system, 5 orbital periods of Jupiter
approximately correspond to 2 periods of Saturn, 13 orbital periods
of Venus approximately correspond to 8 periods of the Earth, and Pluto
makes two orbits for every three of Neptune \citep[cf:][]{Scafetta2014}.
In addition, there are the well-known 1:2:4 resonances of Jupiter's
moons Ganymede, Europa and Io, which was studied by Pierre-Simon Laplace
(1749 -- 1827), and the primary gaps in the asteroid main-belt at
the 4:1, 3:1, 5:2, 7:3, 2:1 mean-motion resonances between the asteroids
and Jupiter, which was first noticed in 1866 by Daniel Kirkwood (1814
-- 1895) \citep{Moons,Moons1998}.

The Trappist-1 solar system is also a very peculiar exoplanetary example.
It is made of a dwarf red star and seven earth-size planets labeled
b, c, d, e, f, g, and h (three are in the habitable zone) whose stable
orbits are characterized by three-body Laplace-type near-resonances
\citep{Gillon,Luger,Tamayo}. The seven orbital periods are: 1.511,
2.422, 4.049, 6.101, 9.207, 12.352, and 18.773 days, respectively
\citep{Agol}. Thus, the period ratios between adjacent planet pairs
(c/b, d/c, e/d, f/e, g/f and h/g) are 1.603, 1.672, 1.507, 1.509,
1.342, 1.520, respectively. These ratios are very close to the following
integer ratios: 8:5, 5:3, 3:2, 3:2, 4:3, 3:2, respectively, with a
mean relative error of 0.6\%, which is the longest known series of
near-resonant exoplanets. The planetary resonances of the Trappist-1
system's motion are so accurate and peculiar that they were translated
into music \citep{Chang,Russo}.

On a wider perspective, gravitational self-organization and harmonic
resonances, and more specifically the planetary invariant inequalities
involving planetary conjunctions and their beats, generate complex
planetary synchronization structures in the solar system that appear
to modulate also solar variability and climate change on Earth. These
phenomena are currently under study by several authors \citep[e.g.:][and others]{Beer,Charvatova,Scafetta2014b,Scafetta2016,Scafetta2020,Stefani,Tattersall}.

The philosophical concept of orbital resonance is known as ``\emph{Musica
Universalis}'' or ``\emph{Music of the Spheres}'' or ``\emph{Harmony
of the Spheres}'', and was first developed in the 6th century BC
by Pythagoras of Samos (570--495 BC) and his followers \citep{Stephenson1974,Godwin,Rogers},
who related planetary periods with the principles of musical harmony.
The philosopher noted that the pitch of a musical note is in inverse
proportion to the length of the string that produces it, and that
intervals between harmonious sound frequencies form simple numerical
ratios. Furthermore, Pythagoras proposed that the bodies of the solar
system (including the Sun, the Moon and the planets) all emit a unique
hum based on their orbital revolution. According to Philolaus (470
-- 385 BC), the planetary harmonics were characterized by four basic
musical intervals: 2:1 (octave), 3:2 (fifth), 4:3 (fourth) and 1:1
(unison).

Herein we adopt a similar transdisciplinary approach and show that
music theory can still be useful to explore some possible unknown
features that characterize the interplanetary gravitational organization
of our solar system. Indeed, Kepler himself was inspired by musical
principles \citep{Cartwright}.

In fact, the correspondence between whole number ratios in orbital
resonances and music theory was further developed by Johannes Kepler
(1571 -- 1630) in \emph{Harmonices Mundi} (The Harmony of the World,
1619), in which he related musical tones with the periods, distances
and angular velocities of the planets \citep[cf.][]{Rogers}. Very
likely, Kepler's conception of the ``Music of the Worlds'' reflected
the polyphony of his day as developed by composers such as Giovanni
Pierluigi da Palestrina (1525 -- 1594).

For example, he noted that, relative to the Sun, the angular speed
of the Earth varies by a semitone (a ratio of 16:15), between aphelion
and perihelion. Similar musical relations were found for the other
planets as well so that Kepler hypothesized the existence of a \emph{celestial
choir} made up of a tenor (Mars), two basses (Saturn and Jupiter),
a soprano (Mercury), and two altos (Venus and Earth). These inquiries
brought forth his discovery of the ``third law of planetary motion''.
We recall that Kepler's first and second law of planetary motions
were proposed 10 years earlier in \emph{Astronomia Nova }(1609). Kepler's
laws were empirically based and published almost 100 years before
Newton proposed the gravitational law that provided their physical
basis.

The third law establishes that the square of a planet's orbital period
$T$ is proportional to the cube of the length of the semi-major axis
of its orbit $a$ as: 
\begin{equation}
a^{3}=cT^{2}=\frac{G(M_{\odot}+m)}{4\pi^{2}}T^{2}\approx\frac{GM_{\odot}}{4\pi^{2}}T^{2},\label{eq1}
\end{equation}
where $G$ is the universal gravitational constant, $m$ is the mass
of the planet and $M_{\odot}$ is the mass of the Sun. Since $M_{\odot}$
is much larger than any planetary mass, $c$ can be considered constant
for the entire solar system. Furthermore, $c=1$ if the period is
measured in years and the semi-major axis length is measured in astronomical
units (that is the mean distance between the Sun and the Earth). Finally,
by establishing a simple relation between the period and the semi-major
axis of an orbit ($a=T^{2/3}$ or $T=a^{3/2}$), Eq. \ref{eq1} allows
the rewriting of any planetary equation depending on one of these
orbital parameters as a function of the other.

Further attempts to model the solar system using simple relations
included the ``Titius--Bode law of Planetary Distances'' \citep{Bode,Titius}
that, with a good approximation, correctly reproduced the orbital
position of Mercury, Venus, Earth, Mars, Jupiter and Saturn, and successfully
predicted those of Ceres and Uranus, although it failed for Neptune.
Additional attempts to improve such methods were proposed by other
authors \citep[e.g.: ][]{Basano,Louise,Molchanov,Nieto}, up to very
recent times \citep[e.g.:][and cited references]{Tattersall,Scafetta2014,Aschwanden}.

\begin{figure}[!t]
\includegraphics[width=1\columnwidth]{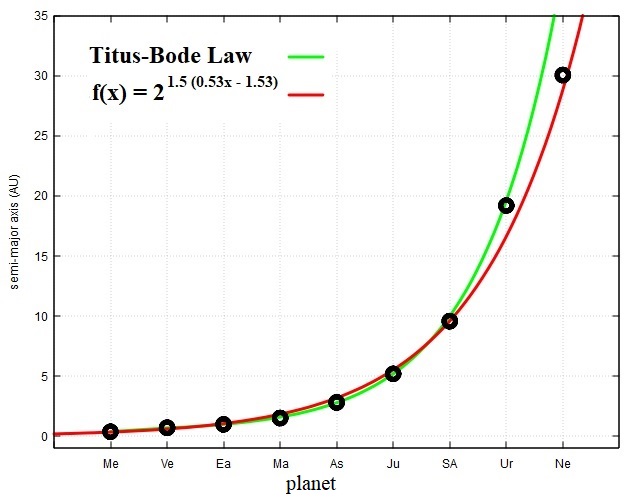}\caption{The semi-major axes of the planets versus the Titus-Bode law and a
simple exponential fit.}
\label{fig1}
\end{figure}

Figure \ref{fig1} shows the semi-major axes of the planets versus
the Titus-Bode empirical law --- $a_{z}=0.4+0.3\times2^{z}$ for
$z=-\infty$ (Mercury), 0 (Venus), 1 (Earth), 2 (Mars), ... , 7 (Neptune)
\citep{Bode,Titius} --- and a simple exponential fit of the type
$f(x)=2^{(ax+b)\times3/2}$, where the integer values of x denote
the planet series number from 1 to 9. The fit gives $a=0.529\pm0.016$
and $b=-1.53\pm0.09$. These coefficients are very close to the whole-ratio
1/2 and -3/2, which might suggest an ideal equation of the type $a_{n}=2^{-9/4}2^{3n/4}\approx0.21\times1.68^{n}$,
where $n$ goes from 1 (Mercury) to 9 (Neptune), and $n=5$ denotes
the asteroid belt.

Similar relations are found in the theoretical literature. For example,
\citet{Pakter} studied the stability and self-organization of planetary
systems similar to our solar system. Gravitational instabilities usually
lead to catastrophic events because planets either collide or are
ejected from the planetary system. However, these authors showed that
if the planetary motion is quasi-periodic and the planets could gain
or lose energy from interplanetary orbiting dust, then, computer simulations
over astronomical time scales suggest that such systems could reach
a planetary self-organized structure. The proposed equations were
the following:

\begin{equation}
m\ddot{x}_{i}=-\frac{GMmx_{i}}{r_{i}^{3}}-\sum_{j}\frac{Gm^{2}(x_{i}-x_{j})}{r_{ij}}-f_{i}^{\theta}\frac{y_{i}}{r_{i}}
\end{equation}

\begin{equation}
m\ddot{y}_{i}=-\frac{GMmy_{i}}{r_{i}^{3}}-\sum_{j}\frac{Gm^{2}(y_{i}-y_{j})}{r_{ij}}-f_{i}^{\theta}\frac{x_{i}}{r_{i}}
\end{equation}
where $M$ is the sun's mass, $m$ is the planet mass (which this
model supposes to all be equal), $r_{i}$ is the distance between
the sun and the planet $i$, $r_{ij}$ is the distance between the
planets $i$ and $j$, and $f_{i}^{\theta}$ is the angular force
responsible for the interaction between the planet $i$ and the residual
dust of the protoplanetary disk.

\citet{Pakter} demonstrated that systems up to 9 planets reach a
self-organized dynamical state where the anomalistic periods between
the radially adjacent planets could be synchronized in a near 2:1
resonance. Moreover, several simulations showed that the ratio of
semi-major axis could follow a geometric progression of the type $r_{n}\sim c^{n}$,
where $c$ is close to 1.6-1.7, which is similar to what was found
for the solar system: see Figure \ref{fig1}. For example, for a perfect
2:1 resonance ratio, the semi-major axis lengths would follow a Titus-Bode-like
relation of the type: $r_{n}\sim2^{2n/3}\approx1.59^{n}$. However,
these authors were not able to find a stable planetary system made
of more than 6 planets. They also acknowledged that the proposed mechanism
could not be ``unique'' in explaining the self-organization of a
solar system. In fact, they did not succeed in exactly simulating
our solar system, which is made of 8 planets of different masses plus
the Asteroid and Kuiper belts.

We also observe that a major problem with the Titius--Bode law (and
also with the above fit function) is that such an equation is physically
unconstrained because the parameter \emph{z} (or \emph{n} in the fitting
function) do not have an upper theoretical limit, which implies that
the equations would eventually fail the prediction. Also using $z=-\infty$
for Mercury in the original Titius--Bode law was arbitrary because
such a value was chosen just to fix the divergence. Therefore, such
equations do not appear to be statistically nor physically robust.
In the following, we propose a mirror-like planetary model that does
not suffer the same problem because it appears to be physically constrained
by the properties of the solar system itself.

In this work, we first extend some of the calculations by \citet{Kepler1619}
to all eight planets of the solar system (in the 17th century the
asteroid belt, Uranus and Neptune were unknown) and, in most cases,
we found that simple whole-number ratios emerge in the periods, distances
and angular velocities only for some planet pairs. This result, however,
implies that a correspondent musical ratio can not be found for all
planetary ratios, which suggests that our solar system is not characterized
by a self-organization structure similar to that found, for example,
in the Trappist-1 solar system. Thus, we searched alternative orbital
metrics and checked whether they could produce a musical correspondence
for all planetary pairs. The assumption is that ratios adopted in
the traditional musical tuning systems and, in particular, those that
form consonances, are peculiar because they are harmoniously interrelated
and, therefore, may unearth important physical relations \citep{Cartwright}.

An interesting feature of the solar system is that it is made of four
inner terrestrial planets (Mercury, Venus, Earth and Mars) and four
outer gas-giant planets (Jupiter, Saturn, Uranus and Neptune) divided
by the asteroid belt plus a large number of asteroids and comets:
see the orbital map of the solar system, art-work by \citet{Lutz}.
Thus, we focused on the mirror symmetries among the distances of the
planets found by \citet{Geddes}, which are today better known only
in the popular scientific literature \citep[e.g.:][]{Martineau}.

These authors noted that the distances of the eight planets of the
solar system from the Sun could be treated as a mirror-reflected system
relative to the belt of the asteroids. Herein, we apply a non-linear
transformation of these equations, which was inspired by the Western
musical practice of dividing the octave (corresponding to a frequency
doubling) into 12 equal parts, called \emph{half-steps}.

These 12 equal parts correspond, for example, to the 12 keys of the
octave of a piano: the correspondent twelve tones are $C$, $Db$,
$D$, $Eb$, $E$, $F$, $Gb$, $G$, $Ab$, $A$, $Bb$, and $B$,
with the $b$ (flat) notes represented by the black keys. Each tone
corresponds to a specific number between 1 and 2 representing a frequency
ratio between that tone and the original reference tone as summarized
in Table \ref{tab:1}. We use the listed musical tones as labels to
express such numerical values.

\begin{table*}[!t]
\centering{}%
\begin{tabular}{|c|l|c|l|r|l|r|c|}
\hline 
\# & Name & Tone & 12-TET & ET-MC & 12-TJI & JI-MC & Consonance\tabularnewline
\hline 
0 & Unison & $C$ & $2^{0/12}=1$ & 0 & $1/1=1$ & 0 & Yes\tabularnewline
\hline 
1 & Minor Second & $Db$ & $2^{1/12}\approx1.0595$ & 100 & $16/15=1.0667$ & 112 & No\tabularnewline
\hline 
2 & Major Second & $D$ & $2^{2/12}\approx1.1225$ & 200 & $9/8=1.125$ & 204 & No\tabularnewline
\hline 
3 & Minor Third & $Eb$ & $2^{3/12}\approx1.1892$ & 300 & $6/5=1.2$ & 316 & Yes\tabularnewline
\hline 
4 & Major Third & $E$ & $2^{4/12}\approx1.2599$ & 400 & $5/4=1.25$ & 386 & Yes\tabularnewline
\hline 
5 & Perfect Fourth & $F$ & $2^{5/12}\approx1.3348$ & 500 & $4/3=1.3333$ & 498 & Yes\tabularnewline
\hline 
6 & Tritone & $Gb$ & $2^{6/12}\approx1.4142$ & 600 & $45/32=1.4063$ & 590 & No\tabularnewline
\hline 
7 & Perfect Fifth & $G$ & $2^{7/12}\approx1.4983$ & 700 & $3/2=1.5$ & 702 & Yes\tabularnewline
\hline 
8 & Minor Sixth & $Ab$ & $2^{8/12}\approx1.5874$ & 800 & $8/5=1.6$ & 814 & Yes\tabularnewline
\hline 
9 & Major Sixth & $A$ & $2^{9/12}\approx1.6818$ & 900 & $5/3=1.6667$ & 884 & Yes\tabularnewline
\hline 
10 & Minor Seventh & $Bb$ & $2^{10/12}\approx1.7818$ & 1000 & $9/5=1.8$ & 1018 & No\tabularnewline
\hline 
11 & Major Seventh & $B$ & $2^{11/12}\approx1.8877$ & 1100 & $15/8=1.875$ & 1088 & No\tabularnewline
\hline 
12 & Octave & $C$ & $2^{12/12}=2$ & 1200 & $2/1=2$ & 1200 & Yes\tabularnewline
\hline 
\end{tabular}\caption{Numerical values of the musical tones (in the key of $C$) of the
equally tempered 12-TET and the justly tuned 12-TJI systems; their
musical cents (MC) values evaluated with Eq. \ref{eq3}; and whether
or not they form a consonance.}
\label{tab:1}
\end{table*}

Of the 12 possible ratios within the octave, only 7 are considered
traditional harmonic consonances \citep{Stephenson1974}. In the key
of $C$, these are labeled $C$, $Eb$, $E$, $F$, $G$, $Ab$ and
$A$, where $C$ is the reference note to which all other tones are
compared. In music, tone pairs are considered harmonic consonances
if they are perceived as ``pleasing'' when sounded together \citep{Thompson}.
Their pleasing quality is thought to result from simple frequency
ratios between their members, namely if the ratios are made of small
whole numbers related to arithmetic and harmonic means. In physics,
consonant ratios could be related to a concept of mutual stability
and balance while dissonant ratios could express some form of instability
or tension.

Inspired by the Classical musical tuning systems, we explore an alternative
way to rewrite the equations proposed by \citet{Geddes} in a very
compact and elegant form, which suggests a possible rational gravitational
organization of the solar system that involves scaling and mirror-symmetries.
The same equations imply ratios by pairs of neighboring planets corresponding
to four main harmonic musical consonances. Our proposed model is then
compared against a recently proposed alternative harmonic orbital
resonance model \citep{Aschwanden} to test its performance in predicting
the positions of the planets of the solar system and found to perform
better. Finally, we respond to the brief, but in our opinion inadequate
critique of \citet{Abhyankar}, which might have prevented until now
a further scientific development of the ideas proposed by \citet{Geddes}
as desired by the same authors.

\section{The 12-TJI and 12-TET tuning systems and their consonances}

In this section, we introduce the reader to some basic concepts of
the music tuning systems, which form the mathematical metric that
we adopt for obtaining our results.

In current Western musical practice, the octave (corresponding to
a doubling of frequency) is almost exclusively divided into 12 parts,
labeled half-steps. The division of the octave into twelve half-steps
likely derives from Pythagorean philosophy in which new tones were
generated by taking the ratio $3/2$. After 12 such iterations, the
pitch ratio to the original note is $(3/2)^{12}=129.746$. Transposing
this note down 7 octaves (dividing the pitch by $2^{7}=128$), there
is a return to the original tone with only a slight discrepancy of
a $1.364\%$ error (known as the Pythagorean Comma). In this way twelve
distinct tones or pitch classes can be defined with only a slight
margin of error: see \citet{Rubinstein} for additional details. Hence,
the octave on the modern piano keyboard has 12 notes.

\begin{figure}[!t]
\centering{}\includegraphics[width=1\columnwidth]{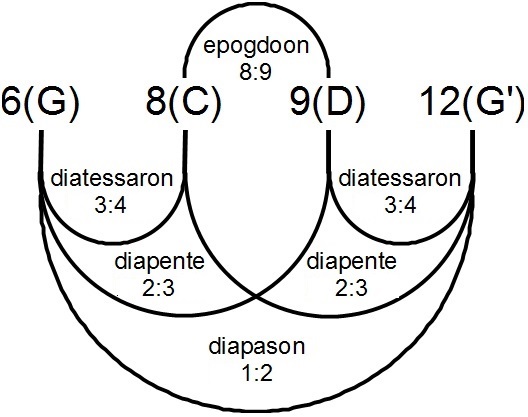}\caption{Pythagorean music theory: diagram showing relations between epogdoon,
diatessaron, diapente, and diapason, which correspond to the Major
Second (9/8), Perfect Fourth (4/3), Perfect Fifth (3/2) and Octave
(2/1), respectively. Pythagorean length ratios in this figure are
the multiplicative inverse of frequency relations as used in this
article.}
\label{figepogdoon}
\end{figure}

The 12-tone system, however, emerged from a long history of evolving
acoustical knowledge and tuning systems. The fundamental ratio of
string lengths is 2 to 1, which is known as the \emph{octave}. The
octave can be divided into two intervals by taking two different means
(for $a=1$ and $b=2$): 
\begin{itemize}
\item arithmetic mean = (a + b) / 2 = 3/2 (known as the ``Perfect Fifth''); 
\item harmonic mean = 2ab / (a + b) = 4/3 (known as the ``Perfect Fourth''). 
\end{itemize}
This yields the four main notes of the Pythagorean musical set: Unison
(1/1), Perfect Fourth (4/3), Perfect Fifth (3/2) and Octave (2/1).
The difference between the Perfect Fifth and the Perfect Fourth gives
the Pythagorean \emph{epogdoon} 9/8 (Major Second), which corresponds
musically to a whole tone, for example the interval from C to D. We
note that the geometric mean ($\sqrt{ab}$) was not used for the calculation
of the harmonic intervals because it produces $\sqrt{2}$, which is
an irrational number that the Pythagoreans considered imperfect. Also
the number 17 was considered imperfect because it separates the 16
from its \emph{epogdoon} 18. Figure \ref{figepogdoon} summarizes
the Pythagorean music theory. The four notes were discussed in Plato's
Timaeus where they were related to the harmony of the cosmos \citep{Godwin};
they are based on the numbers 1, 2, 3 and 4 (the sum is 10), which
formed the mystical symbol of the \emph{tetractys} that symbolized
the \emph{musica universalis}, the Cosmos, the four classical elements
(fire, air, water, and earth) and the organization of space.

The four notes are also labeled \emph{harmonic consonances} since
they sound ``pleasant'' when played together. However, this property
should not be understood just as a human perception, but as a consequence
of the mathematical simple ratios that characterize these notes so
that the subjective gradation from consonance to dissonance should
correspond to a gradation of sound-frequency ratios from simple ratios
to more complex ones. The International Cyclopedia of Music and Musicians
\citep{Thompson} explains: ``\emph{Acoustically, consonance is the
degrees of blending and fusion between two or more tones. The lower
the ratio, such as 2:1, 3:2, 4:3, the greater the degree of fusion,
hence consonance. Consonance may also be distinguished by the degree
of freedom from beats. The octave (2:1) is the most perfect acoustic
consonance and is free of beats; then follows, in order of degree,
the fifth (3:2), the fourth (4:3), etc.; the series becoming more
and more dissonant as the ratios depart from the simplest, i.e. 2:1.}''

Several scientific works have linked the acoustics of consonances
to their aesthetic qualities \citep{Bones,McDermott,Plack}. However,
these ratios appear to have also a deep geometrical and physical meaning,
which may explain why they may be relevant to reveal some hidden organization
of the solar system and of other physical systems. For example, the
four main tones of the Pythagorean musical set describe a Keplerian
orbit. In fact, if $a$ is the semi-major axis, $b$ the semi-minor
axis and $l=a(1-e^{2})$, it is found that $a$ is the arithmetic
mean, $b$ the geometric mean and $l$ the harmonic mean of $r_{min}$
and $r_{max}$ \citep{Cartwright}.

In the 16th century, musical theorists such as Gioseffo Zarlino (1517--1590)
completed the traditional set of harmonic consonances by adding four
more intervals: the ``Minor Third'' (6/5); the ``Major Third''
(5/4); the ``Minor Sixth'' (8/5); and the ``Major Sixth'' (5/3)
\citep{Zarlino}. These ratios can also be derived as harmonic and
arithmetic means of 1/1 and 3/2 in the case of thirds and means between
4/3 and 2/1 in the case of sixths \citep{Forster}. The proposed system
was defined as \emph{just} because all notes are related by intervals
that are defined by rational numbers \citep{Cartwright}. In the tuning
systems employing just-intonation, exact harmonic consonance ratios
are maximized.

In our analysis, we employ the five-limit twelve-tone scale which
maximizes just intonation between tone pairs of the octave; we will
refer to this as the twelve-tone just-intonation system (12-TJI).
In this system, more complex ratios are assigned to the five remaining
non-consonant (or dissonant) tones of the octave: (in the key of C)
the ``Minor Second'' ($Db$, 16/15); the ``Major Second'' ($D$,
9/8); the ``Tritone'' ($Gb$, 45/32); the ``Minor Seventh'' ($Bd$,
9/5); and the ``Major Seventh'' ($B$, 15/8).

The utilization of perfect ratios of harmonic consonances within just-intonation
tuning causes the size of half-steps to vary between different pairs
of adjacent notes within the octave, as shown below. As a consequence,
a keyboard instrument tuned to play in one key (e.g. $C$ major) can
be wildly out of tune in another (e.g. $Gb$ major).

In order to minimize this limitation, different tuning systems evolved
through the centuries, eventually yielding the twelve-tone equal-temperament
(12-TET) system which, today, has been widely adopted. In general,
an equal-tempered system is a musical tuning system that approximates
just intervals by dividing an octave into equal steps. Therefore,
the ratio of the frequencies of any adjacent pair of notes is the
same. In the 12-TET, the smallest interval is a 1/12th of the width
of an octave, and it is called a \emph{semitone} or \emph{half-step}.
By normalizing the ratio of all half-steps in the octave to a value
of 
\begin{equation}
\gamma=2^{1/12}\approx1.05946,\label{eq2}
\end{equation}
one can play equally in tune in any key (e.g. $C$ major, $Gb$ major,
etc.), but this is achieved at the expense of introducing slight errors
into all the perfect simple ratios of harmonic consonances, except
the Unison and the Octave. A 12-TET system would have been unimaginable
to ancient Greek philosophers because it involved roots of 2 which
are irrational numbers.

To evaluate quantitatively how well specific tone pairs are tuned
in the standard 12-TET system versus those in the 12-TJI system, the
octave (that is the 2/1 ratio) is assumed to be made of 1200 cents.
Given a real number $x$ between 1 and 2, its musical cent (MC) value
is defined by the equation: 
\begin{equation}
MC(x)=1200\log_{2}(x).\label{eq3}
\end{equation}

Thus, for $r=1$ (Unison) its MC value is 0 cents, while for $r=2$
(Octave) its MC value is 1200 cents. Moving sequentially by $n$ equal-tempered
half-steps from a reference tone to a new tone, the frequency ratio
between the two tones is exactly $\gamma^{n}=2^{n/12}$ so that its
MC value is $100\times n$. If $r$ is larger than 2, Eq. \ref{eq3}
can still be used, and the correct tone relation can be estimated
by subtracting an integer number of 1200 cents for each octave. For
example, $MC=1200+100\times n$ cents ($x=2^{1+n/12}$) corresponds
to the $n^{th}$ equal-tempered tone in the second octave; $MC=2400+100\times n$
cents ($x=2^{2+n/12}$) corresponds to the $n^{th}$ equal-tempered
tone in the third octave; and so on.

In the idealized 12-TJI system, which uses the whole number ratios
of harmonic consonances, the tones slightly diverge from the 12-TET
ones: for example, a Perfect Fifth, a 3/2 ratio, has 702 cents while
the 7 (equal-tempered) half-steps approximates a Perfect Fifth with
$2^{7/12}$, which corresponds to 700 cents; similarly, a Major Third
is 5/4 and corresponds to 386 cents, whereas 4 equal-tempered half-steps
correspond to 400 cents; and so on. The differences between the 12-TJI
and the 12-TET vary from 0 to 18 cents, with an average of 9 cents.
Table \ref{tab:1} summarizes and compares the 12 tones in the 12-TET
and 12-TJI systems.

The same MC notation can also be used to evaluate how close the ratios
of orbital parameters between adjacent planets are to those of musical
tones pairs in the 12-TJI and 12-TET systems.

In the following we will use both tuning systems: the 12-TJI allows
a direct interpretation based on harmonic whole number ratios that
recalls the resonance formalism of the orbital commensurabilities;
the 12-TET system allows us to write the equations in a more compact
mathematical formalism.

\section{Basic attempts to find musical tones and consonances in the orbital
parameter ratios of adjacent planets}

In \emph{Harmonices Mundi }\citet{Kepler1619} attempted to find correspondences
between ratios of planetary orbital parameters, musical harmony, and
Platonic solids. Herein, we evaluate and study the ratios of the semi-major
axis, sidereal periods, and mean orbital velocities of neighboring
planets, including the asteroid belt.

The analysis aims to determine whether musical tones and, more specifically,
consonances exist in such a series of orbital variables as found,
for example, in the Trappist-1 solar system or among the moons of
Jupiter. Table \ref{tab:2} reports the astronomical data herein used.
The planets are listed from the closest to the farther from the Sun
as: Mercury (Me); Venus (Ve); Earth (Ea); Mars (Ma); Asteroid (As);
Jupiter (Ju); Saturn (Sa); Uranus (Ur); and Neptune (Ne).

\begin{table}[!t]
\centering{}%
\begin{tabular}{|c|r|r|r|r|}
\hline 
Planet & S-M Axis & Period & Speed & $a^{2/3}$\tabularnewline
 & AU & year & km/s & \tabularnewline
\hline 
\hline 
Mercury & 0.387 & 0.241 & 47.36 & 0.531\tabularnewline
\hline 
Venus & 0.723 & 0.615 & 35.02 & 0.806\tabularnewline
\hline 
Earth & 1.000 & 1.000 & 29.78 & 1.000\tabularnewline
\hline 
Mars & 1.524 & 1.881 & 24.07 & 1.324\tabularnewline
\hline 
Asteroid & 2.825 & {*} 4.748 & {*} 17.72 & 1.998\tabularnewline
\hline 
Jupiter & 5.204 & 11.862 & 13.06 & 3.003\tabularnewline
\hline 
Saturn & 9.583 & 29.457 & 9.68 & 4.512\tabularnewline
\hline 
Uranus & 19.201 & 84.011 & 6.80 & 7.170\tabularnewline
\hline 
Neptune & 30.048 & 164.79 & 5.43 & 9.665\tabularnewline
\hline 
\end{tabular}\caption{The semi-major axes $a$, sidereal period $T$ and mean orbital velocities
$v$ of the planets of the solar system, including the belt of the
asteroids located between Mars and Jupiter. Asteroid refers to the
5:2 Kirkwood asteroid-belt gap; ({*}) indicates theoretical values
calculated using Eq. \ref{eq1}. The last column reports the values
of $a^{2/3}$ discussed in the main text. (Data from: \protect\href{https://nssdc.gsfc.nasa.gov/planetary/factsheet/}{https://nssdc.gsfc.nasa.gov/planetary/factsheet/}).}
\label{tab:2}
\end{table}

The asteroid belt distance from the Sun (As) was set at the 5:2 Kirkwood-gap
($2.825$ AU) as a surrogate (although negative or missing) planet
because, in the following discussion, such a region is supposed to
be a kind of ``divergence'' or ``reflection'' point separating
the inner (or terrestrial) from the outer (or gas-giant) planets \citep[cf.: ][]{Geddes,Moons}.
This is the central gap of the asteroid belt and it is linked to the
5:2 resonance between Jupiter and Saturn. Such a distance is very
close to the geometrical mean between Mars and Jupiter, $\sqrt{Ma\times Ju}\approx2.816$
AU, and to the mean orbital radius of the dwarf planet Ceres (about
$2.769$ AU).

\begin{table*}[!t]
\centering{}%
\begin{tabular}{|c|c|r|lc|r|lc|r|}
\hline 
Planets & Ratio & MC & 12-TET & (Con) & ET-MC\textsubscript{err} & 12-TJI & (Con) & JI-MC\textsubscript{err}\tabularnewline
\hline 
\hline 
\multicolumn{9}{|c|}{Semi-Major Axis Length}\tabularnewline
\hline 
Ve/Me & 1.868 & 1082 & $B$ & (N) & -18 & $B$ & (N) & -6\tabularnewline
\hline 
Ea/Ve & 1.383 & 562 & $Gb$ & (N) & -38 & $Gb$ & (N) & -28\tabularnewline
\hline 
Ma/Ea & 1.524 & 729 & $G$ & (Y) & 29 & $G$ & (Y) & 27\tabularnewline
\hline 
As/Ma & 1.854 & 1068 & $B$ & (N) & -32 & $B$ & (N) & -20\tabularnewline
\hline 
Ju/As & 1.842 & 1058 & $B$ & (N) & -42 & $B$ & (N) & -30\tabularnewline
\hline 
Sa/Ju & 1.841 & 1057 & $B$ & (N) & -43 & $Bb$/$B$ & (N) & 31/-31\tabularnewline
\hline 
Ur/Sa & 2.004 & 1203 & $C$ & (Y) & 3 & $C$ & (Y) & 3\tabularnewline
\hline 
Ne/Ur & 1.565 & 775 & $Ab$ & (Y) & -25 & $Ab$ & (Y) & -39\tabularnewline
\hline 
\multicolumn{9}{|c|}{Orbital Period}\tabularnewline
\hline 
Ve/Me & 2.552 & {*} 422 & $E$ & (Y) & 22 & $E$ & (Y) & 36\tabularnewline
\hline 
Ea/Ve & 1.626 & 842 & $Ab$ & (Y) & 42 & $Ab$ & (Y) & 28\tabularnewline
\hline 
Ma/Ea & 1.881 & 1094 & $B$ & (N) & -6 & $B$ & (N) & 6\tabularnewline
\hline 
As/Ma & 2.524 & {*} 403 & $E$ & (Y) & 3 & $E$ & (Y) & 17\tabularnewline
\hline 
Ju/As & 2.498 & {*} 385 & $E$ & (Y) & -15 & $E$ & (Y) & -1\tabularnewline
\hline 
Sa/Ju & 2.484 & {*} 375 & $E$ & (Y) & -25 & $E$ & (Y) & -11\tabularnewline
\hline 
Ur/Sa & 2.852 & {*} 614 & $Gb$ & (N) & 14 & $Gb$ & (N) & 24\tabularnewline
\hline 
Ne/Ur & 1.962 & 1166 & $C$ & (Y) & -34 & $C$ & (Y) & -34\tabularnewline
\hline 
\multicolumn{9}{|c|}{Average Orbital Speed}\tabularnewline
\hline 
Me/Ve & 1.352 & 523 & $F$ & (Y) & 23 & $F$ & (Y) & 25\tabularnewline
\hline 
Ve/Ea & 1.176 & 281 & $Eb$ & (Y) & -19 & $Eb$ & (Y) & -35\tabularnewline
\hline 
Ea/Ma & 1.237 & 369 & $E$ & (Y) & -31 & $E$ & (Y) & -17\tabularnewline
\hline 
Ma/As & 1.358 & 530 & $F$ & (Y) & 30 & $F$ & (Y) & 32\tabularnewline
\hline 
As/Ju & 1.357 & 529 & $F$ & (Y) & 29 & $F$ & (Y) & 31\tabularnewline
\hline 
Ju/Sa & 1.349 & 518 & $F$ & (Y) & 18 & $F$ & (Y) & 20\tabularnewline
\hline 
Sa/Ur & 1.424 & 611 & $Gb$ & (N) & 11 & $Gb$ & (N) & 21\tabularnewline
\hline 
Ur/Ne & 1.252 & 389 & $E$ & (Y) & -11 & $E$ & (Y) & 3\tabularnewline
\hline 
\end{tabular}\caption{Musical tones in the semi-major axis, orbital period, average speed
ratios of adjacent planets. The columns indicate: the adjacent planet
couples; the orbital parameter ratio; its value in musical cents (MC),
the symbol ({*}) indicates that the MC value is reduced by 1200 cents;
12-TET is the closest tone (for the inner planet of the pair when
the outer planet is tuned to $C$) according to the exact 12-TET system
with the Yes/No consonance property (Con); ET-MC\protect\protect\textsubscript{err}
is the relative error; 12-TJI and JI-MC\protect\protect\textsubscript{err}
are equivalent but relative to the 12 tones of the just-intonation
12-TJI system. (Refer to Table \ref{tab:1} for the numerical values
of the tones).}
\label{tab:3}
\end{table*}

Table \ref{tab:3} shows the chosen orbital ratios, their values in
MC (using Eq. \ref{eq3}), and their closest tone (using Table \ref{tab:1})
relative to both the 12-TET and the 12-TJI systems. In addition, we
report in parenthesis whether or not the closest tone is a consonance
and the error-distance in MC of the planetary ratio from the closest
tone for both musical systems. Here we maintain Kepler's practice
of assigning the higher frequency tone to the faster moving, inner
planet. We also assume that when the discrepancy between the orbital
ratio and the music tone is equal or larger than 25 cents, the two
values are not compatible (let us say ``untuned'') and, therefore,
such an orbital ratio cannot be interpreted as a musical tone in these
tuning systems.

Table \ref{tab:3} shows that out of 8 planetary adjacent couples: 
\begin{itemize}
\item \textbf{using the semi-major axis --} 6 or 5 ratios, in 12-TET and
12-TJI respectively, are untuned, and only 3 are close to consonances,
of which 2 are untuned. 
\item \textbf{using the orbital period -- }3 ratios are untuned, and only
6 are close to consonances, of which 3 are untuned. 
\item \textbf{using the average speed --} 3 or 4 ratios are untuned, and
only 7 are close to consonances, of which 3 or 4 are untuned, respectively. 
\end{itemize}
Thus, the results are not satisfactory as each of these three metrics
fails to find tuned musical correlates for several planetary pair
ratios. However, one could wonder whether a different set of orbital
measures could provide a better musical interpretation of the movements
of the bodies of the solar system.

In this regard, we notice that Kepler's third law of planetary motions
(Eq. \ref{eq1}) indicates the existence of simple relations between
the ratios of planetary measurements, which are characterized by specific
exponents such as 
\begin{equation}
\left\{ \begin{array}{lll}
(a_{1}/a_{2})^{3/2} & = & T_{1}/T_{2}\\
(a_{1}/a_{2})^{-1/2} & = & v_{1}/v_{2},
\end{array}\right.\label{eq4}
\end{equation}
where the first equation derives directly from Kepler's third law
and the second by approximating the orbital perimeter as $2\pi a$,
where $a$ is the semi-major axis, and using the definition of mean
orbital speed as $v=2\pi a/T$. Thus, the ratios of average distances,
orbital periods, and average velocities are mutually related by varying
the exponents from $1$, to $3/2$, to $-1/2$ respectively.

Thus, in the following sections, we look for exponents $k$ such that
the values 
\begin{equation}
(a_{1}/a_{2})^{k}\label{eq5}
\end{equation}
for adjacent planets could be best expressed in musical tones and,
more specifically, in harmonic consonances. To do this, we take advantage
of the equations proposed by \citet{Geddes}.

\section{The Geddes -- King-Hele equations}

\citet{Geddes} noted that the orbits of the eight planets of the
solar system appear as if they were ``reflected'' about the asteroid
belt so that the following symmetries are found: Mercury$\leftrightarrow$Neptune,
Venus$\leftrightarrow$Uranus, Earth$\leftrightarrow$Saturn, and
Mars$\leftrightarrow$Jupiter.

More specifically, these authors found that the ratios among the mean
distances from the Sun of the planets (in the following denoted by
the planet's name initials) could be approximated as powers of a single
constant which these authors denoted by 
\begin{equation}
r=2^{1/8}\approx1.09051.\label{eq6}
\end{equation}
Then, the following nearly exact equations relating contiguous planets
were found: 
\begin{equation}
\left\{ \begin{array}{lll}
Ve\approx r^{7}Me; & Ea\approx r^{4}Ve; & Ma\approx r^{5}Ea;\\
As\approx r^{7}Ma; & Ju\approx r^{7}As;\\
Sa\approx r^{7}Ju; & Ur\approx r^{8}Sa; & Ne\approx r^{5}Ur.
\end{array}\right.\label{eq7}
\end{equation}

The middle equations relate Mars, an estimate of the asteroid belt
distance from the Sun and Jupiter, where the distance of the asteroid
belt ($As$) was originally arbitrarily set at $\sqrt{Ma\times Ju}\approx2.816$
AU, which is, nevertheless, very close to the 5:2 Kirkwood-gap at
$2.825$ AU that we prefer to use in the following as the mirror point.

\citet{Geddes} noted that the percent errors in the eight equations
listed in the system \ref{eq7} are very small. We get: 1.9\%, -2.2\%,
-1.2\%, 0.8\%, 0.8\%, 0.4\%, 0.2\%, 1.5\% respectively.

The above equations can also be combined in several ways. For example,
it is possible to obtain the mean distance from the Sun of all planets
as a function of only that of Mercury and specific powers of $r$.
It is also easy to obtain the following identity: 
\begin{equation}
\frac{Ve\times Ju}{Me\times Sa}\approx\frac{Ea\times Ne}{Ma\times Ur}\approx1,\label{eq8}
\end{equation}
which have an accuracy of 1.5\% and 2.7\%, respectively. Finally,
it is possible to obtain the following Geddes -- King-Hele equations
that ``mirror'' the planets relative to the asteroid belt: 
\begin{equation}
\frac{Me}{Ea}\times\frac{Ne}{Sa}\approx\frac{Ve}{Me}\times\frac{Ur}{Ne}\approx\frac{Ea}{Ma}\times\frac{Sa}{Ju}\approx r^{2}=2^{1/4}\label{eq9}
\end{equation}
\begin{equation}
\frac{Sa}{Ea}\times\frac{Ma}{Ju}\approx\frac{Ur}{Ve}\times\frac{Ea}{Sa}\approx\frac{Ne}{Me}\times\frac{Ve}{Ur}\approx r^{12}=2^{3/2}\label{eq10}
\end{equation}

It is to be noted that the chosen constant $r$ was interpreted by
the authors as the mean frequency ratio between notes in a musical
octave, although \citet{Abhyankar} critiqued such an interpretation.
However, as explained above, in Western musical practice, the octave
is divided into 12 parts, not 8.

\section{A non-linear transformation of the Geddes -- King-Hele equations}

Herein we convert the Geddes -- King-Hele equations to a form that
is compatible with the chromatic musical scale. This is done by raising
each side of Eqs. \ref{eq7} to the $k=2/3$ power. In fact, as previously
stated, the frequency ratio between any adjacent half-steps in the
12-TET system is $\gamma=2^{1/12}\approx1.05946$ (Eq. \ref{eq2}),
which is equal to $r^{2/3}$.

\citet{Abhyankar} also noted that such a change of metric would make
the Geddes -- King-Hele equations more compatible with the frequency
ratios between the notes of the tuning systems proposed above. However,
he summarily and erroneously concluded that there was ``nothing particularly
musical'' about the mirror-symmetries between the distances of the
planets nor that those symmetries were ``telling us something about
the origin of the Solar System or its stability''. Indeed, he did
not realize the mathematical and physical properties of the new equations
that the new metric implies. Let us disclose it.

The new planetary terms can now be related mathematically by powers
of $\gamma$, which is equivalent to movements in half-steps in the
12-TET musical system because $\gamma=r^{2/3}$. Therefore, it follows
that: 
\begin{equation}
\left\{ \begin{array}{lll}
Ve'\approx\gamma^{7}Me'; & Ea'\approx\gamma{}^{4}Ve'; & Ma'\approx\gamma{}^{5}Ea';\\
As'\approx\gamma{}^{7}Ma'; & Ju'\approx\gamma{}^{7}As';\\
Sa'\approx\gamma{}^{7}Ju'; & Ur'\approx\gamma{}^{8}Sa'; & Ne'\approx\gamma{}^{5}Ur'.
\end{array}\right.\label{eq11}
\end{equation}
where $Me'=Me^{2/3}$, $Ve'=Ve^{2/3}$, $Ea'=Ea^{2/3}$, $Ma'=Ma^{2/3}$,
$As'=As^{2/3}$, $Ju'=Ju^{2/3}$, $Sa'=Sa^{2/3}$, $Ur'=Ur^{2/3}$and
$Ne'=Ne^{2/3}$.

We can now employ the 12-TET system to express the ratios between
adjacent planets in terms of musical half-steps. For example, $Ve'=\gamma^{7}Me'$
can be rewritten as: 
\begin{equation}
\left(\frac{Ve}{Me}\right)^{\frac{2}{3}}=\frac{Ve'}{Me'}\approx\gamma^{7}=2^{7/12}.\label{eq12}
\end{equation}
This is equivalent to saying that the ratio of semi-major axis lengths
of the orbits of Venus and Mars elevated to the 2/3rd power, that
is $Ve'/Me'$, is equal to the ratio of frequencies of two pitches
that are 7 half-steps apart, which corresponds to a Perfect Fifth
(a ratio of 3/2).

Eq. \ref{eq10} can be rewritten as: 
\begin{equation}
\frac{Sa'}{Ea'}\times\frac{Ma'}{Ju'}\approx\frac{Ur'}{Ve'}\times\frac{Ea'}{Sa'}\approx\frac{Ne'}{Me'}\times\frac{Ve'}{Ur'}\approx\gamma^{12}=2,\label{eq13}
\end{equation}
which expresses octave ratios.

\citet{Abhyankar} was able to derive Eq. \ref{eq13}, but he did
not realize that it implies a series of scaling musical relations.
In fact: $Ju'/Ma'\approx\gamma^{14}$ = 14 half-steps $\approx$ 2
Perfect Fifths = $\left(3/2\right)\times\left(3/2\right)=2.25$; $Sa'/Ea'$
adds an octave to that, doubling the ratio to 4.5; similarly, $Ur'/Ve'$
doubles this to 9; and $Ne'/Me'$ doubles that to 18. Consequently,
Eq. \ref{eq13} can be rewritten also in the following compact form:
\begin{equation}
1\frac{Ne'}{Me'}\approx2\frac{Ur'}{Ve'}\approx4\frac{Sa'}{Ea'}\approx8\frac{Ju'}{Ma'}\approx18,\label{eq14}
\end{equation}
which reveals, both a mirror-like and scaling structure relative to
the asteroid belt. Note the sequence of the powers of 2 ($1=2^{0}$,
$2=2^{1}$, $4=2^{2}$ and $8=2^{3}$) relative to planetary pairs
approaching the asteroid belt.

As a function of the semi-major axis $a$, of the period $T$ and
of the mean orbital speed $v$, Eq. \ref{eq14} corresponds to \begin{strip} \begin{eqnarray} 1\left(\frac{a_{Ne}}{a_{Me}}\right)^{\frac{2}{3}}\approx2\left(\frac{a_{Ur}}{a_{Ve}}\right)^{\frac{2}{3}}\approx4\left(\frac{a_{Sa}}{a_{Ea}}\right)^{\frac{2}{3}}\approx8\left(\frac{a_{Ju}}{a_{Ma}}\right)^{\frac{2}{3}} & \approx & 18\label{eq15}\\ 1\left(\frac{T_{Ne}}{T_{Me}}\right)^{\frac{4}{9}}\approx2\left(\frac{T_{Ur}}{T_{Ve}}\right)^{\frac{4}{9}}\approx4\left(\frac{T_{Sa}}{T_{Ea}}\right)^{\frac{4}{9}}\approx8\left(\frac{T_{Ju}}{T_{Ma}}\right)^{\frac{4}{9}} & \approx & 18\label{eq16}\\ 1\left(\frac{v_{Me}}{v_{Ne}}\right)^{\frac{4}{3}}\approx2\left(\frac{v_{Ve}}{v_{Ur}}\right)^{\frac{4}{3}}\approx4\left(\frac{v_{Ea}}{v_{Sa}}\right)^{\frac{4}{3}}\approx8\left(\frac{v_{Ma}}{v_{Ju}}\right)^{\frac{4}{3}} & \approx & 18\label{eq17} \end{eqnarray} \end{strip}where
the Eqs. \ref{eq4} were used. Using Table \ref{tab:2}, the exact
values of the four ratios multiplied by increasing powers of 2 are:
18.20, 17.80, 18.05 and 18.14, respectively, using the semi-major
orbital axes; 18.20, 17.79, 17.99 and 18.14, respectively, using the
periods; and 17.95, 17.79, 17.90 and 18.08, respectively, using the
mean orbital velocities. Thus, Eq. \ref{eq14} {[}or Eqs. \ref{eq15},
\ref{eq16} and \ref{eq17}{]} describes the orbits of the planets
of the solar system within about $1\%$ error (or with a 99\% accuracy),
and elegantly expresses its scaling and mirror-like symmetry structure
with respect to the asteroid belt. More sprecifically, it suggests
that the inner and outer orbits of the solar system are organized
in a simple scaling structure.

Using the equations for planetary distances raised to the 2/3rd power
and expressing the results in terms of half-steps, we now evaluate
how well the planetary ratios can be assigned to musical tones.

\begin{table*}[!t]
\centering{}%
\begin{tabular}{|c|c|c|c|c|lll|c|c|c|c|}
\hline 
Planets & Ratio & Ratio$^{2/3}$ & MC & NHS & \multicolumn{3}{c}{Tone and Consonance} & ET-MC & ET-MC\textsubscript{err} & JI-MC & JI-MC\textsubscript{err}\tabularnewline
\hline 
Ve/Me & 1.868 & 1.517 & 721 & 7 & $G$ & Perfect Fifth & (Y) & 700 & 21 & 702 ($3/2$) & 19\tabularnewline
\hline 
Ea/Ve & 1.383 & 1.241 & 374 & 4 & $E$ & Major Third & (Y) & 400 & -26 & 386 ($5/4$) & -12\tabularnewline
\hline 
Ma/Ea & 1.524 & 1.324 & 486 & 5 & $F$ & Perfect Fourth & (Y) & 500 & -14 & 498 ($4/3$) & -12\tabularnewline
\hline 
As/Ma & 1.854 & 1.509 & 712 & 7 & $G$ & Perfect Fifth & (Y) & 700 & 12 & 702 ($3/2$) & 10\tabularnewline
\hline 
Ju/As & 1.842 & 1.503 & 705 & 7 & $G$ & Perfect Fifth & (Y) & 700 & 5 & 702 ($3/2$) & 3\tabularnewline
\hline 
Sa/Ju & 1.841 & 1.502 & 705 & 7 & $G$ & Perfect Fifth & (Y) & 700 & 5 & 702 ($3/2$) & 3\tabularnewline
\hline 
Ur/Sa & 2.004 & 1.589 & 802 & 8 & $Ab$ & Minor sixth & (Y) & 800 & 2 & 814 ($8/5$) & -12\tabularnewline
\hline 
Ne/Ur & 1.565 & 1.348 & 517 & 5 & $F$ & Perfect Fourth & (Y) & 500 & 17 & 498 ($4/3$) & 19\tabularnewline
\hline 
\end{tabular}\caption{Number of half-steps (NHS), musical tones and consonances in both
the exact 12-TET (ET) and 12-TJI (JI) values, and in musical cents
(MC), versus the semi-major axis ratios of adjacent planets elevated
to the 2/3rd power, with the relative musical cent errors (MC\protect\protect\textsubscript{err}).
(Refer to Table \ref{tab:1} for the numerical values of the tones).}
\label{tab:4}
\end{table*}

\begin{figure*}[!t]
\centering{}\includegraphics[width=1\textwidth]{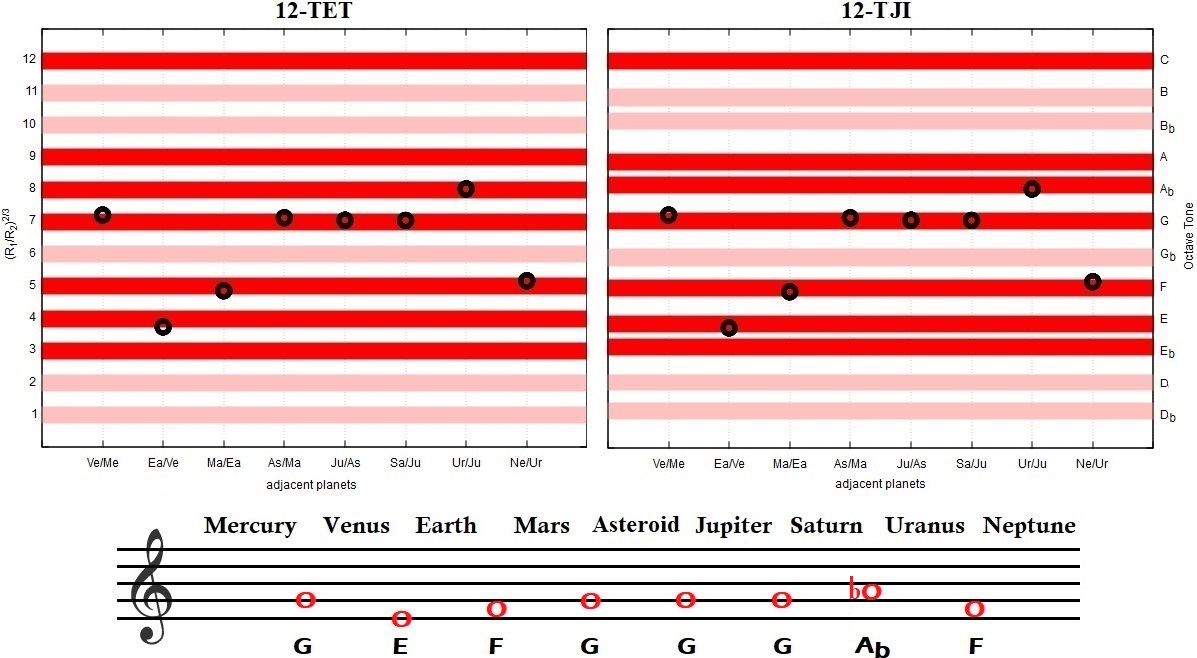}\caption{Top: The black circles indicate the distance ratios raised to the
2/3rds power of adjacent planets in both the equal tempered 12-TET
and justly tuned 12-TJI (in the key of $C$ expressed as half-steps
or MC/100): see Table \ref{tab:4}. The red bars indicate the consonances,
while the pink ones the dissonances. The width of each bar is 50 cents.
Bottom: the musical notation gives the note for the inner planet of
the pair when the outer planet is tuned to middle $C$. (Refer to
Table \ref{tab:1} for the numerical values of the tones).}
\label{fig2}
\end{figure*}

Table \ref{tab:4} reports the semi-major axis ratios of adjacent
planets elevated to the 2/3rd power using the orbital data listed
in Table \ref{tab:2}, and compares them with their closest musical
tones using both the 12-TET and 12-TJI values (expressed in musical
cents, using Eq. \ref{eq3}). We also tabulated the relative musical
cent errors from the closest tone. Figure \ref{fig2} shows the results
using both tone systems.

Using the 12-TET system, an absolute divergence between 2 and 26 cents,
with an average of 13 cents is observed; whereas in the 12-TJI tuning,
the absolute divergence is between 3 and 19 cents, with an average
of 11 cents. These results suggest that the chosen planetary measure
among couples of adjacent planets of the solar system can be well
approximated by the tones of the Western musical tradition.

As previously stated, of the 12 possible tones within the octave,
7 are considered consonances \citep{Stephenson1974}. This happens
when the exact value of the note can be well approximated by a ratio
$n/m$ where the whole numbers $n$ and $m$ are small and members
of the set: 3, 5, and powers of 2. The consonance ratios, in the key
of $C$, are: Unison or Octave ($C$), 1/1 or 2/1; Minor Third ($Eb$),
6/5; Major Third ($E$), 5/4; Perfect Fourth ($F$), 4/3; Perfect
Fifth ($G$), 3/2; Minor Sixth ($Ab$), 8/5; Major Sixth ($A$), 5/3.
As summarized in Table \ref{tab:5}, $Ve'/Me'$, $As'/Ma'$, $Ju'/As'$,
and $Sa'/Ju'$ are tuned to a Perfect Fifth, $Ma'/Ea'$ and $Ne'/Ur'$
are tuned to a Perfect Fourth, $Ea'/Ve'$ is tuned to a Major Third,
and $Ur'/Sa'$ to a Minor Sixth.

\begin{table*}[!t]
\centering{}%
\begin{tabular}{|c|l|c|c|c|}
\hline 
\# & Name & Tone & Consonance & Planet ratio\tabularnewline
\hline 
1 & Minor second & $Db$ & No & \tabularnewline
\hline 
2 & Major second & $D$ & No & \tabularnewline
\hline 
3 & Minor third & $Eb$ & Yes & \tabularnewline
\hline 
4 & Major Third & $E$ & Yes & $Ea'/Ve'$\tabularnewline
\hline 
5 & Perfect Fourth & $F$ & Yes & $Ma'/Ea'$; $Ne'/Ur'$\tabularnewline
\hline 
6 & Tritone & $Gb$ & No & \tabularnewline
\hline 
\multirow{2}{*}{7} & \multirow{2}{*}{Perfect Fifth} & \multirow{2}{*}{$G$} & \multirow{2}{*}{Yes} & $Ve'/Me'$; $As'/Ma'$;\tabularnewline
 &  &  &  & $Ju'/As'$; $Sa'/Ju'$\tabularnewline
\hline 
8 & Minor sixth & $Ab$ & Yes & $Ur'/Sa'$\tabularnewline
\hline 
9 & Major Sixth & $A$ & Yes & \tabularnewline
\hline 
10 & Minor seventh & $Bb$ & No & \tabularnewline
\hline 
11 & Major Seventh & $B$ & No & \tabularnewline
\hline 
12 & Octave & $C$ & Yes & \tabularnewline
\hline 
\end{tabular}\caption{Comparison between the 12 tones and the semimajor axis ratios among
the adjacent planets raised to the $k=2/3$ power deduced from the
corrected Geddes--King-Hele planetary distance equations (Eq. \ref{eq11}).
The tone is that of the inner planet of the pair when the outer planet
is tuned to C. (Refer to Table \ref{tab:1} for the numerical values
of the tones).}
\label{tab:5}
\end{table*}

Thus, not only do all these planetary ratios appear sufficiently well-tuned
to be members of the traditional music octave, but all of them also
correspond to the musical consonance ratios according to both the
12-TET and 12-TJI systems. This result suggests that, taken as a set,
the orbits of the planets present a specific well tuned and harmonized
structure.

\section{Statistical significance and robustness of the exponent k = 2/3}

We now check whether the close fit between idealized musical ratios
and planetary data using the exponent $k=2/3$ is coincidental. Thus,
we repeated the above calculations by varying the value of $k$ between
0.3 to 1 in steps of 0.001.

\begin{figure}[!t]
\centering{}\includegraphics[width=1\columnwidth]{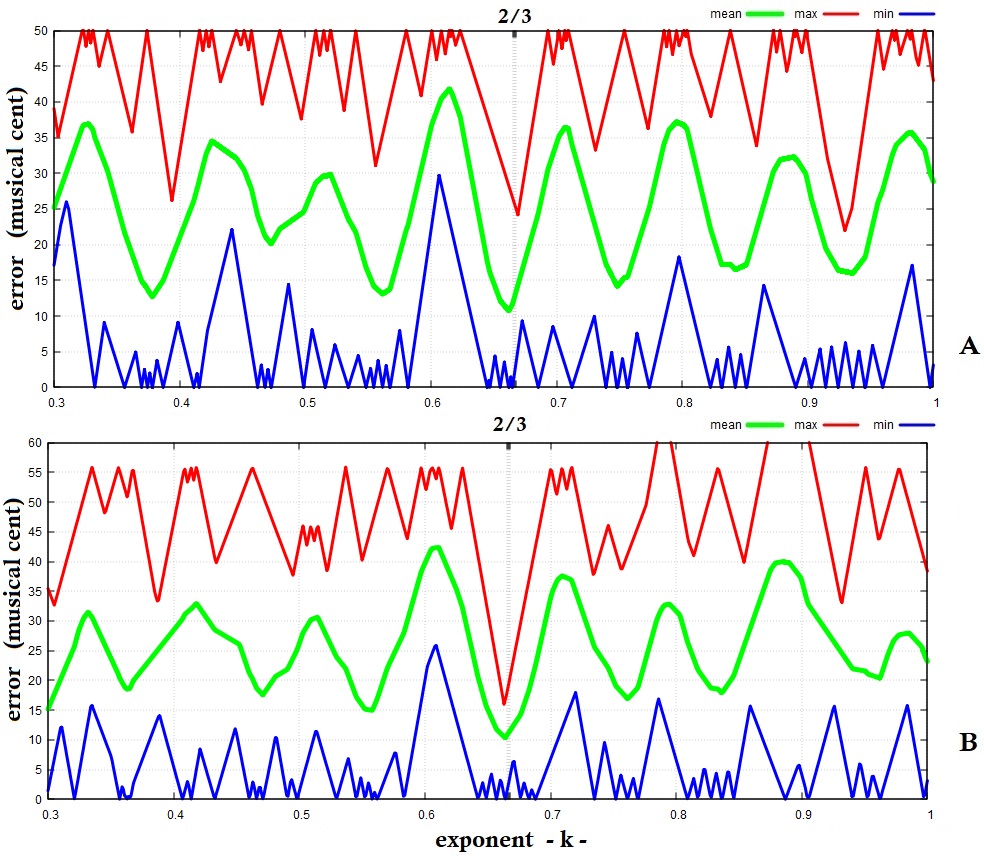}\caption{Exponent $k$ versus average, maximum and minimum error measured in
musical cents from adjacent planet distance ratios raised to the exponent
$k$ to the closest musical tone for: A) the equal tempered 12-TET
system; B) the justly tuned 12-TJI. The value for $k=2/3\approx0.667$
is highlighted.}
\label{fig3}
\end{figure}

\begin{figure}[t]
\centering{}\includegraphics[width=1\columnwidth]{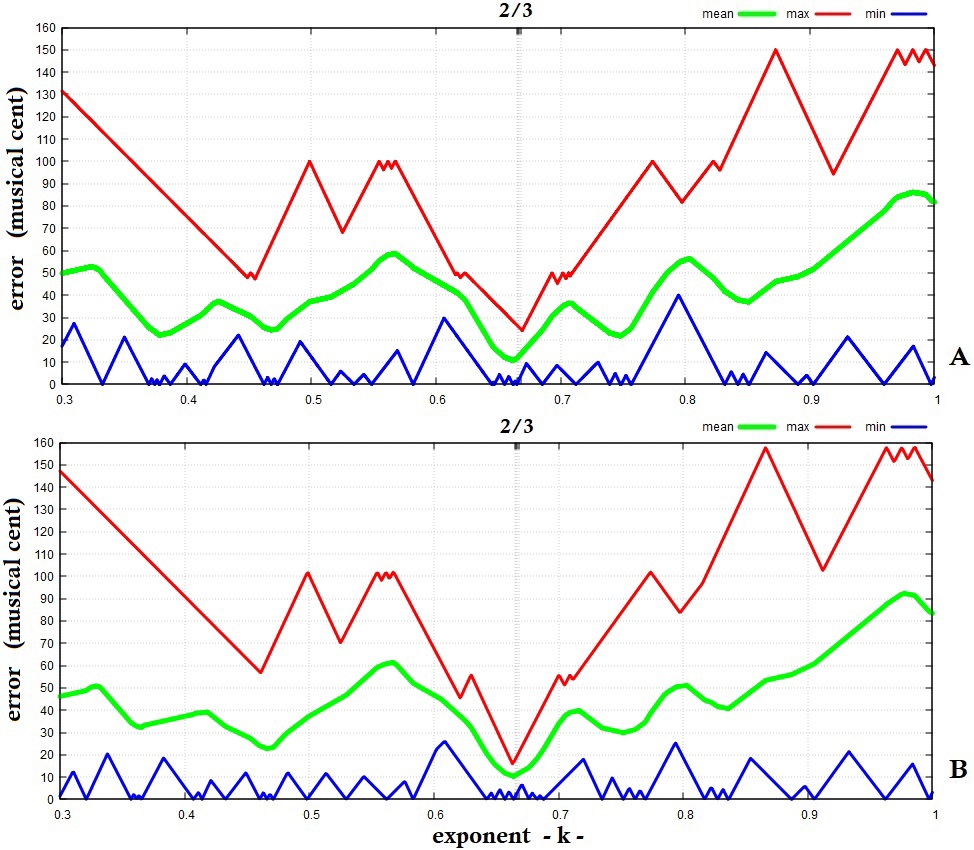}\caption{Exponent $k$ versus average, maximum and minimum error measured in
musical cents from adjacent planet distance ratios raised to the exponent
$k$ to the closest musical consonance for: A) 12-TET system; B) the
12-TJI system. The value for $k=2/3\approx0.667$ is highlighted.}
\label{fig4}
\end{figure}

Figure \ref{fig3} plots the average, maximum and minimum error (measured
in MC) between the eight adjacent planet semi-major axis ratios raised
to the power of $k$ and the closest of the twelve musical tones listed
in Table \ref{tab:1} as a function of the exponent $k$. Figure \ref{fig3}A
uses the tones of the 12-TET system while Figure \ref{fig3}B uses
those of the 12-TJI system. The analysis depicted in Figure \ref{fig4}
uses only the 7 consonances and ignores the other 5 tones. The position
of the value for $k=2/3$ is highlighted in both figures.

Figures \ref{fig3} and \ref{fig4} show that $k\approx2/3$ corresponds
to the absolute minimum in the average error (green curve) between
our proposed model and both musical tuning systems, which suggests
that the chosen measure could be physically meaningful. Regarding
the maximal error (red curve), $k\approx2/3$ corresponds in Figure
\ref{fig3}A to the second-lowest minimum, and to the absolute minimum
in Figure \ref{fig3}B and in both Figures \ref{fig4}A and \ref{fig4}B,
which refer to the consonances alone. The result suggests that the
exponent $k=2/3$ optimizes the metrics expressed by Eq. \ref{eq5}
to reproduce the simple ratios found in musical consonances.

To further test the statistical relevance of our result, we now evaluate
the probability of obtaining the 8 planetary ratios close to musical
consonances against the null hypothesis that they are randomly distributed.
Above we found that the planetary data yield the simple ratios of
harmonic consonances within a mean accuracy of about 11-13 cents up
to a maximum of 26 cents in just a single case relative to the 12-TET
system. The range in MC of planetary distance ratios elevated to the
2/3rd power is 428 cents: from 374 cents (close to a Major Third for
Ea'/Ve') to 802 cents (close to a Minor Sixth for Ur'/Sa'). To each
end of this range, we add the associated error-value, to obtain a
total band range of 475 cents from 350 to 825 cents. Within this range,
four idealized consonance ratios lie: the Major Third, the Perfect
Fourth, the Perfect Fifth, and the Minor Sixth. Each of these four
tones can be assumed to have a maximum error range of $\pm$25 cents.
Therefore, their total error range accommodates at most 200 cents
of the total available 475 cent range, that is 50 cents for each of
the four consonances. Thus, the statistical chance that the eight
planetary ratios occur randomly with this proximity to the four selected
consonances is $p=(200/475)^{8}<0.001=0.1\%$. Thus, it is very unlikely
that our result occurs by chance.

\section{Comparison versus the harmonic orbit resonance model}

\citet{Aschwanden} studied the regularity of the spaced patterns
of the distances of the planets of the solar system and showed that
logarithmic spacing models, such as both the Titus-Bode law and its
generalized form, perform poorly versus a harmonic resonance model
based on quantized scaling factors. More specifically, the author
showed that the planet distances $R_{i}$ from the Sun and their orbital
periods $T_{i}$ (where $i=1$ for Mercury, $i=2$ for Venus, etc.)
are related to scaling laws of the type: 
\begin{equation}
\frac{R_{i+1}}{R_{i}}=\left(\frac{T_{i+1}}{T_{i}}\right)^{\frac{2}{3}}=\left(\frac{H_{i+1}}{H_{i}}\right)^{\frac{2}{3}}\label{eq:18.1}
\end{equation}
where the $2/3$ exponent derives from Eq. \ref{eq1}, and the whole
number ratios $H_{i+1}/H_{i}$ yield to the following planetary equations
linking the semi-major axes of neighboring planet pairs: 
\begin{equation}
\left\{ \begin{array}{ll}
Ve\approx(5/2)^{2/3}Me; & Ea\approx(5/3)^{2/3}Ve;\\
Ma\approx(2/1)^{2/3}Ea; & As\approx(5/2)^{2/3}Ma;\\
Ju\approx(5/2)^{2/3}As; & Sa\approx(5/2)^{2/3}Ju;\\
Ur'\approx(3/1)^{2/3}Sa; & Ne\approx(2/1)^{2/3}Ur.
\end{array}\right.\label{eq19-1}
\end{equation}

The Eqs. \ref{eq19-1} express a planetary model of the solar system
alternative to that of Eqs. \ref{eq11}. The scaling factors are different.
Thus, it is important to determine which one of the two planetary
models performs better in predicting the size of the orbits of the
solar system.

To do this, we observe that given a semi-major axis $a$ for a planet
(see Table \ref{tab:2}), the equation sets \ref{eq11} and \ref{eq19-1}
can be used to predict the position of the other eight planets relative
to the chosen one. Thus, for each of the two models, we can obtain
nine different sets of predictions starting from each planet. Finally,
the two prediction groups are statistically compared.

\begin{figure*}[!t]
\centering{}\includegraphics[width=1\textwidth]{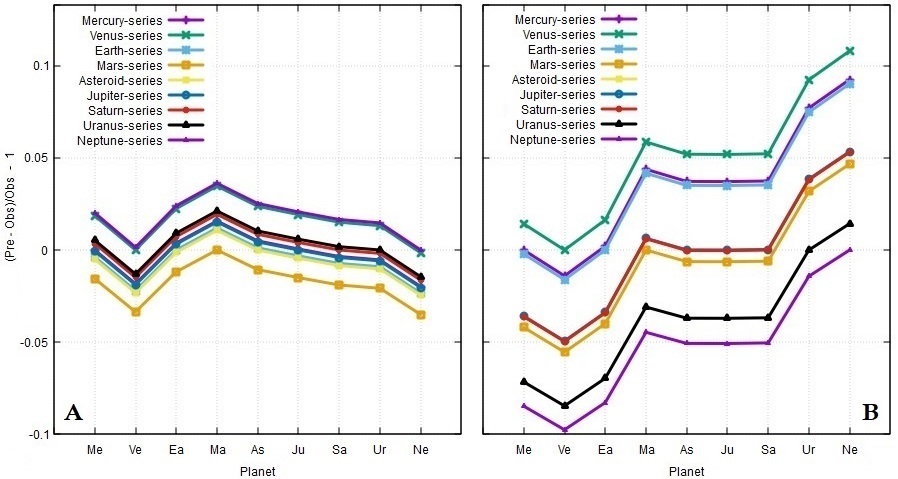}\caption{Relative errors regarding the predictions of the semi-major axis lengths
of the planets of the solar system according to: {[}A{]} the Geddes
-- King-Hele 12-TET based model (Eqs. \ref{eq11}) herein proposed;
{[}B{]} the harmonic orbit resonance model by \citet{Aschwanden}
(Eqs. \ref{eq19-1}).}
\label{fig5}
\end{figure*}

This is done in Figure \ref{fig5} that shows the relative error for
each planet given by the expression: 
\begin{equation}
relative\;error=\frac{prediction-observation}{observation}-1.
\end{equation}
The figure shows that the Geddes -- King-Hele 12-TET based model
(Eqs. \ref{eq11}) performs significantly better than the harmonic
orbit resonance model by \citet{Aschwanden} (Eqs. \ref{eq19-1}).
This is demonstrated by the lower dispersion and trend bias in the
relative error sets produced by the former model relative to the latter
one. Table \ref{tab:6} reports the average predictions for each planet
relative to the nine sets for each model with the relative errors
relative to the observations. It is found that on average the Geddes
-- King-Hele 12-TET model predicts the correct semi-major axis lengths
with a mean error of 0.8\%, while the harmonic orbit resonance model
has a mean error of 2.5\%. The latter also presents a trend bias because
the semi-major axis lengths of Mercury, Venus and Earth are underestimated
by about 4\% while those of Uranus and Neptune are overestimated by
about 5\%.

\begin{table*}[!t]
\centering{}%
\begin{tabular}{|c|c|c|c|c|c|}
\hline 
Planet & S-M Axis & \multicolumn{2}{c|}{GKH 12-TET} & \multicolumn{2}{c|}{HOR}\tabularnewline
\hline 
 & Observations & Predictions & Error (\%) & Predictions & Error (\%)\tabularnewline
\hline 
Mercury & 0.387 & 0.388 & 0.26 & 0.374 & -3.26\tabularnewline
\hline 
Venus & 0.723 & 0.712 & -1.58 & 0.690 & -4.62\tabularnewline
\hline 
Earth & 1.000 & 1.006 & 0.63 & 0.969 & -3.06\tabularnewline
\hline 
Mars & 1.524 & 1.552 & 1.83 & 1.539 & 0.97\tabularnewline
\hline 
Asteroid & 2.825 & 2.846 & 0.75 & 2.834 & 0.34\tabularnewline
\hline 
Jupiter & 5.204 & 5.220 & 0.31 & 5.221 & 0.33\tabularnewline
\hline 
Saturn & 9.583 & 9.574 & -0.10 & 9.617 & 0.36\tabularnewline
\hline 
Uranus & 19.201 & 19.147 & -0.28 & 20.005 & 4.19\tabularnewline
\hline 
Neptune & 30.048 & 29.529 & -1.73 & 31.756 & 5.68\tabularnewline
\hline 
\end{tabular}\caption{Mean predictions and relative errors in the semi-major axes of the
planets of the solar system relative to: {[}A{]} the Geddes -- King-Hele
12-TET based model (Eqs. \ref{eq11}) herein proposed; {[}B{]} the
harmonic orbit resonance (HOR) model by \citet{Aschwanden} (Eqs.
\ref{eq19-1}).}
\label{tab:6}
\end{table*}

The poor performance of the harmonic orbit resonance model was somewhat
expected from the results discussed in section 3 because it approximates
the orbital period ratios listed in Table \ref{tab:3} with consonances
(within two octaves), which, however, are not always the closest of
the twelve tones to the period ratios. Even if the harmonic orbit
resonance model would have adopted the more accurate ratios listed
in Table \ref{tab:3}, it would have nevertheless inadequately agreed
with the data, as we discussed in Section 3.

\section{Prediction of the Kirkwood gaps of the asteroid-belt}

As a final test, we extend the Geddes -- King-Hele 12-TET model to
evaluate its prediction ability. For example, the increasing powers
of 2 present in Eq. \ref{eq15} as the planet pairs approach the mirror
point given by the asteroid belt, suggest the existence of a final
step characterized by the multiplicative factor $2^{4}=16$, which
is the highest possible value compatible with the constant 18 that
satisfies the condition $a_{i+1}/a_{i}>1$.

This extension suggests the existence of one last inner pair of close
orbits (semi-major axis $a_{1}<a_{2}$) that are mirror-symmetric
relative to the asteroid belt and that fulfill the following condition:
\begin{equation}
16\left(\frac{a_{2}}{a_{1}}\right)^{\frac{2}{3}}\approx18.\label{eq21}
\end{equation}
Since the two orbits would be very close to each other, they should
characterize the geometry of the asteroid belt itself. Note that the
ratio $18/16=9/8$ corresponds to the Pythagorean \emph{epogdoon}
(Figure \ref{figepogdoon}).

Indeed, the asteroid main-belt is characterized by five primary gaps
at the 4:1, 3:1, 5:2, 7:3, 2:1 mean-motion resonances between the
asteroids and Jupiter \citep{Moons,Moons1998}. The central region
is characterized by the three central gaps at $a_{3:1}=2.502$ AU,
$a_{5:2}=2.825$ AU, and $a_{7:3}=2.958$ AU, respectively. Ceres
is nearly in the middle at $a_{c}=2.769$ AU. By assuming the mirror
point at Ceres or at $a_{5:2}$ (which is what we adopted above for
the entire solar system), we have the possible mirroring pair given
by $a_{3:1}\leftrightarrow a_{7:3}$. By applying Eq. \ref{eq21},
we get: 
\begin{equation}
16\left(\frac{a_{7:3}}{a_{3:1}}\right)^{\frac{2}{3}}=16\left(\frac{3}{7}\times\frac{3}{1}\right)^{\frac{4}{9}}=16\left(\frac{9}{7}\right)^{\frac{4}{9}}=17.89,\label{eq22}
\end{equation}
where we used Eqs. \ref{eq1} and \ref{eq15}, together with the period
of Jupiter (11.86 years) to get $a_{3:1}$ and $a_{7:3}$. Thus, the
prediction of Eq. \ref{eq21} has an error of 0.6\% and it is linked
to Jupiter's resonances.

It is interesting to notice that if the ratios $As/Ma$ and $Ju/As$
(where above we chose $As=a_{5:2}$ as the mirror point and the asteroid-belt
position) are evaluated as $a_{3:1}/Ma$ and $Ju/a_{7:3}$, we would
get 572 and 652 MC, respectively, which do not correspond to any tone
and occur near 600 MC that corresponds to a Tritone (a dissonant tone
relation). Perhaps, these relations explain why $a_{3:1}$ and $a_{7:3}$
are gaps despite their orbital resonance with Jupiter and, in general,
why the asteroid belt occupies an unstable gravitational region.

In conclusion, Eq. \ref{eq15} with its mathematical extension Eq.
\ref{eq21} appears to well characterize the regularity of the spaced
patterns of the distances of the planets of the solar system including
the inner main structure of the asteroid belt.

\section{Vulcanoid asteroids versus transneptunian objects}

Eq. \ref{eq15} plus Eq. \ref{eq21} should complete our model, which
is, therefore, physically fully constrained. In fact, no other extension
of the equation would be possible since it would require an additional,
but unknown terrestrial planet between the Sun and Mercury --- the
mythical planet Vulcan that Urbain Le Verrier suggested in the 1850s
to explain the anomalies of the orbit of Mercury (a problem that was
later solved by Albert Einstein) or some vulcanoid asteroids which
are still hypothesized \citep{Evans} --- and another planet between
Neptune and the termination shock or the heliopause boundary (between
30 and 100 AU from the Sun) where only small transneptunian objects
like Pluto, Eris, and other comets are found, which are not classified
as regular planets of the solar system. Furthermore, it is not possible
to extend our model beyond the gaps of the asteroid main-belt limit
expressed by Eq. \ref{eq21}.

This property greatly differentiates our model from, for example,
the Titius--Bode's law whose upper limit (and also the lower limit
in the case of Mercury) is unconstrained and, therefore, also yields
questions of statistical robustness.

On the contrary, our equation is fully constrained. Thus, the statistical
robustness of its predictions cannot be easily questioned, and it
establishes that it is possible to evaluate the planetary orbits of
the outer planets from those of the inner planets up to the gaps of
the asteroid-belt with a single scaling and mirror-like equation (depicted
graphically in Figure \ref{fig6}) within an average error of 1\%
or, alternatively, with a 99\% accuracy.

In any case, let us try to extend the proposed model further; an operation
that can be done by doubling and doubling again Eq. \ref{eq15} for
each pair of added symmetric bodies that, in the case of the solar
system, can only generically represent astronomical bands, as actual
real planets are missing.

By assuming Pluto ($a_{P}=39.237$ au), which could represent the
Kuiper belt, and by doubling Eq. \ref{eq15} to accommodate another
planet pair, the semi-major axis of its specular body, the mythical
planet Vulcan, would be expected at $a_{V}\approx39.237/(18*2)^{3/2}=0.182$
au between the Sun and Mercury. Relative to their neighboring planets
--- Neptune and Mercury, respectively --- we have $(a_{P}/Ne)^{2/3}=1.195\approx6/5$
(Minor Third) and $(Me/a_{V})^{2/3}=1.65\approx5/3$ (Major Sixth),
which are both consonances.

By assuming Eris ($a_{E}=67.9$ au), which could represent the Scattered
disk (a scarcely populated region at the boundary of the solar system),
and by doubling again Eq. \ref{eq15} to accommodate a second planet
pair, its specular scattered disk would be expected at $a_{Sz}\approx67.9/(18*4)^{3/2}=0.111$
au. Relative to their neighboring planets --- Pluto and Vulcan, respectively
--- we have $(a_{E}/a_{P})^{2/3}=1.44\approx45/32$ (Tritone) and
$(a_{V}/a_{Sz})^{2/3}=1.39\approx45/32$ (Tritone), which are both
dissonant tones, as scattered regions would suggest.

Beyond the Scattered disk there is only the Oort cloud.

Thus, the Kuiper belt and the Scattered disk could be specular to
the hypothesized vulcanoid asteroid belts and gaps, which could theoretically
exist inside the orbit of Mercury at distances of 0.06-0.21 au from
the Sun \citep{Evans}.

With the last considerations, we have completed the description of
the solar system with a single scaling-mirror equation: Eq. \ref{eq15}
plus Eq. \ref{eq21}, plus its possible extensions obtained by doubling
it for each pair of additional specular bodies or gravitational bands.

\section{Discussion}

The dynamics of celestial bodies follow the laws of gravitation complemented
by some dissipative processes. The problem is that, even in the simplest
case of three bodies interacting gravitationally, a closed-form solution
for this case does not exist. However, empirical evidences suggest
that orbital systems can self-organize in alternative synchronization
structures which are not yet fully understood.

In the specific case of the solar system, we found that the rewriting
of the Geddes -- King-Hele equations in the proposed new form yields
a very compact and elegant expression --- Eq. \ref{eq14} or, equivalently,
Eqs. \ref{eq15}, \ref{eq16}, and \ref{eq17} --- which appears
to disclose the hidden gravitational self-organization structure of
our planetary system. When raised to the 2/3 power (a non-linear transformation),
the orbits of the planets show a rational organization that is not
apparent in the non-transformed orbital parameters. The fact that
the 2/3 exponent minimizes the deviations and the planetary equations
are more accurate than alternative harmonic resonance models, supports
the robustness of our result.

\begin{figure*}[!t]
\centering{}\includegraphics[width=1\textwidth]{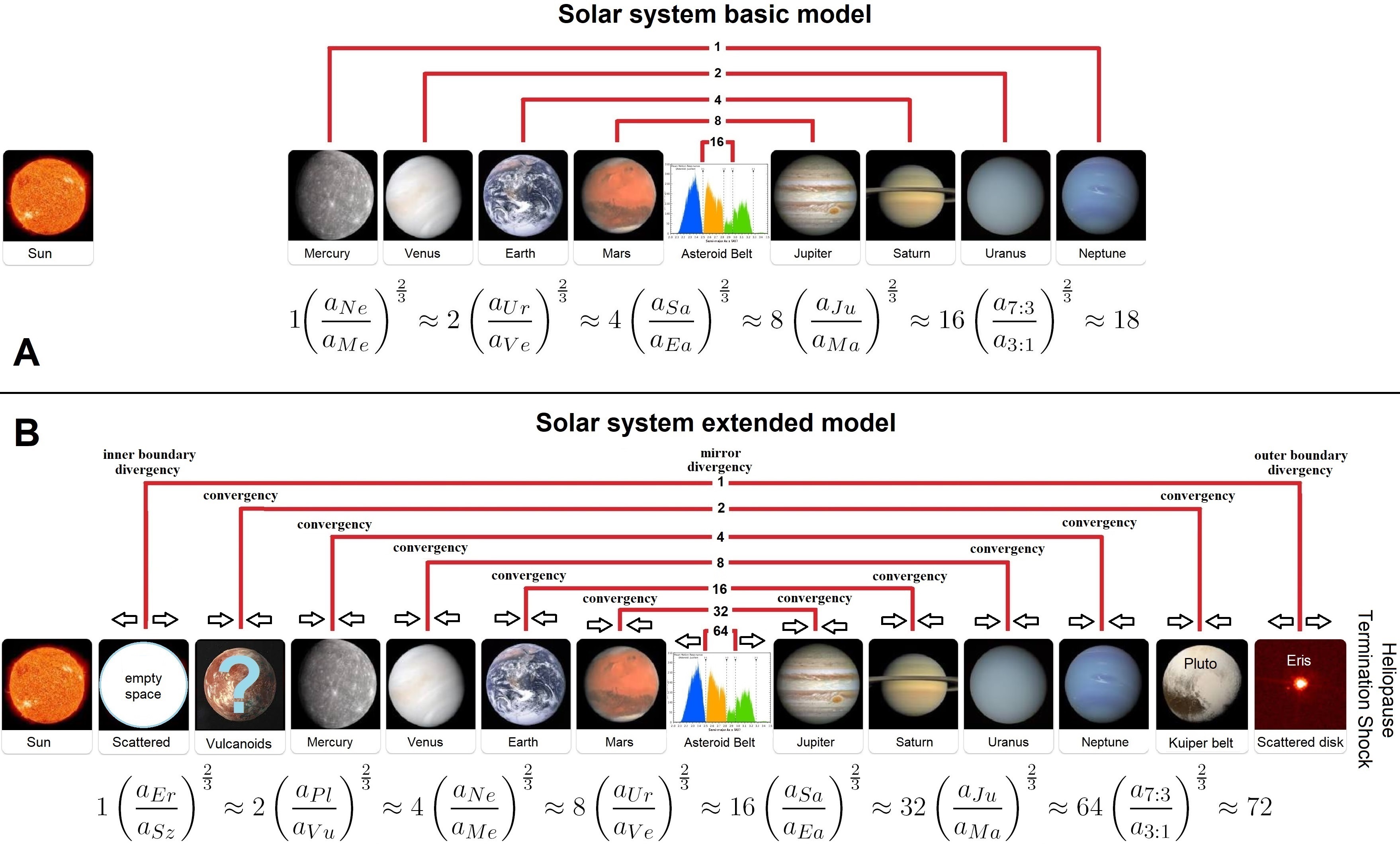}\caption{{[}A{]} Graphical representation of Eq. \ref{eq:25} expressing the
scaling and the mirror symmetries among the semi-major axis lengths
of the eight planets of the solar system, including the central region
of the asteroid belt, according to the 12-TET model of the Geddes
-- King-Hele equations (99\% accuracy). {[}B{]} Graphical representation
of the hypothesized scaling and mirror-symmetric planetary organization
of the solar system according to the extended model (Eq. \ref{eq:26}).
The ratio sequences are based on the Pythagorean \emph{epogdoon} (the
interval ratio $9/8=18/16=72/64$) and its addition with up to six
octaves.}
\label{fig6}
\end{figure*}

Figure \ref{fig6}A graphically represents the orbital scaling and
mirror-symmetries of the solar system in the following compact equation:
\begin{strip} \begin{align} 1\left(\frac{a_{Ne}}{a_{Me}}\right)^{\frac{2}{3}}\approx2\left(\frac{a_{Ur}}{a_{Ve}}\right)^{\frac{2}{3}}\approx4\left(\frac{a_{Sa}}{a_{Ea}}\right)^{\frac{2}{3}}\approx8\left(\frac{a_{Ju}}{a_{Ma}}\right)^{\frac{2}{3}} & \approx16\left(\frac{a_{7:3}}{a_{3:1}}\right)^{\frac{2}{3}}\approx18, \label{eq:25} \end{align} \end{strip}which
links together the eight planets of the solar system plus the asteroid
belt by highlighting the scaling and mirror symmetries among their
semi-major axis lengths according to the 12-TET model of the Geddes
-- King-Hele equations. The above equation has a clear aesthetic
appeal, and its five ratios have an accuracy of 99\%.

Eq. \ref{eq:25} can also be further extended by doubling it for each
pair of additional mirror-symmetric bodies. The first two possible
extensions appear to have a physical meaning because they would correspond
to the bands of the hypothesized vulcanoid asteroids versus the transneptunian
objects and to the internal and external limits of the planetary disk
of the solar system. According to this model, the planetary disk of
the solar system would be constrained between two dissonant regions,
that is a divergent or scattered zone very close to the sun (Sz) at
about 0.1 au, and an equally divergent specular region corresponding
to the Scattered disk (represented by Eris) and extending up to about
100 au from the sun. The asteroid belt, represented by Ceres, at about
2.0-3.5 au divides the planetary disk into an inner and outer region;
this belt would also be a dissonant-divergent zone. Then, the inner
region split into five rings would correspond to the orbits of Mars,
Earth, Venus, Mercury, and, finally, the hypothesized Vulcanoid belt
close to the sun. Similarly, the outer region split into five rings
would correspond to the orbits of Jupiter, Saturn, Uranus, Neptune,
and, finally, the Kuiper belt (represented by Pluto). In this way,
the planetary disk of the Solar System would be fully described by
the following extended mirror-scaling equation: \begin{strip} \begin{align} 1\left(\frac{a_{Er}}{a_{Sz}}\right)^{\frac{2}{3}}\approx2\left(\frac{a_{Pl}}{a_{Vu}}\right)^{\frac{2}{3}}\approx4\left(\frac{a_{Ne}}{a_{Me}}\right)^{\frac{2}{3}}\approx8\left(\frac{a_{Ur}}{a_{Ve}}\right)^{\frac{2}{3}}\approx\nonumber \\ 16\left(\frac{a_{Sa}}{a_{Ea}}\right)^{\frac{2}{3}}\approx32\left(\frac{a_{Ju}}{a_{Ma}}\right)^{\frac{2}{3}} & \approx64\left(\frac{a_{7:3}}{a_{3:1}}\right)^{\frac{2}{3}}\approx72,\label{eq:26} \end{align} \end{strip}which
is shown in Figure \ref{fig6}B. The two sequences suggest that the
orbital scaling-mirror symmetries of the solar system are expressed
by the Pythagorean \emph{epogdoon} (the tone ratio $9/8=18/16=72/64$)
and its addition with one or more octaves. 

Table \ref{tab:7} summaries the value and accuracy of each element
of Eq. \ref{eq:26} obtained with the orbital data in Table \ref{tab:2}.
The equation was divided by 64 for normalization. We found that each
planetary-pair ratio differs from the Pythagorean \emph{epogdoon}
(9/8) by at most 1\%. 

\begin{table*}[!t]
\centering{}%
\begin{tabular}{c|c|c|c|c|c|c|c|c}
Eq. 26 & $\frac{1}{64}\left(\frac{a_{Er}}{a_{Sz}}\right)^{\frac{2}{3}}$ & $\frac{1}{32}\left(\frac{a_{Pl}}{a_{Vu}}\right)^{\frac{2}{3}}$ & $\frac{1}{16}\left(\frac{a_{Ne}}{a_{Me}}\right)^{\frac{2}{3}}$ & $\frac{1}{8}\left(\frac{a_{Ur}}{a_{Ve}}\right)^{\frac{2}{3}}$ & $\frac{1}{4}\left(\frac{a_{Sa}}{a_{Ea}}\right)^{\frac{2}{3}}$ & $\frac{1}{2}\left(\frac{a_{Ju}}{a_{Ma}}\right)^{\frac{2}{3}}$ & $1\left(\frac{a_{7:3}}{a_{3:1}}\right)^{\frac{2}{3}}$ & 9/8\tabularnewline
 &  &  &  &  &  &  &  & \tabularnewline
val. & 1.125 {*} & 1.125 {*} & \textbf{1.137} & \textbf{1.113} & \textbf{1.128} & \textbf{1.134} & \textbf{1.118} & 1.125\tabularnewline
acc. & ( 100\% ) & ( 100\% ) & \textbf{( 101\% )} & \textbf{( 99\% )} & \textbf{( 100\% )} & \textbf{( 101\% )} & \textbf{( 99\% )} & \tabularnewline
\end{tabular}\caption{Value and accuracy of each element of Eq. \ref{eq:26} obtained with
the orbital data in Table \ref{tab:2}. ( {*} Theoretical).}
\label{tab:7}
\end{table*}

Both Eqs. \ref{eq:25} and \ref{eq:26} can be rewritten as functions
of the orbital periods or mean speeds by substituting the exponent
2/3 with 4/9 and 4/3 as in Eqs. \ref{eq16} and \ref{eq17}, respectively.
This transformation does not allow us to interpret the quotients as
simple orbital period ratios. Nor does it allow us to do so for orbital
frequency relations, which are simply the multiplicative inverses
of those derived from orbital periods. The traditional understanding
of a gravitational self-organization structure involves some form
of linear relation among the orbital frequencies and excludes the
application of non-linear transformations. Nevertheless, Eqs. \ref{eq:25}
and \ref{eq:26} express relations of pure rational numbers as all
quotients are non-dimensional, which is a strong indication of some
kind of (still unknown) synchronization phenomena that can emerge
from planetary dynamics produced by adding more gravitational bodies
and dissipative interactions (tidal forces, radiation pressure, friction,
etc.). The question remains on how a non-linear transformation of
orbital periods forms these rational ratios, and this represents a
challenge from a dynamical point of view that can be addressed in
future research. Nevertheless, the robustness of the results is impressive.
Once the exponent of the non-linear transformation is optimized at
$k=2/3$ (as depicted in Figures \ref{fig3} and \ref{fig4}), the
transformed orbital ratios between neighboring planetary pairs are
compatible not only with the tone relations of the traditional 12-TJI
and 12-TET musical tuning systems, but also specifically with their
consonances. The fact that the error is slightly smaller for just
intonation (12-TJI) compared with equal tempered (12-TET) tuning may
suggest that the solar system is mostly characterized by super-particular
ratios, that is by 3/2, 4/3, 5/4 and 6/5 (including Pluto) used in
the former. Moreover, Eqs. \ref{eq:25} and \ref{eq:26} are based
on the super-particular ratio 9/8 (Major Second) multiplied by powers
of 2. The ratio 9 to 8 was known in Pythagorean music theory as the
\emph{epogdoon}, which corresponds to the whole tone and is derived
from the Pythagorean consonances. By extending the mirror-scaling
equation (Eq. \ref{eq:26}) by two additional rings and noting the
planetary contiguous ratios produced, we see that all musical tone
relations from the Major Second to the Major Sixth are represented
in this model.

The physical interpretation of the result is still based on preliminary
planetary models, analogies and speculations.

For example, \citet{Pakter} proposed a planetary model suggesting
that, under specific constraints and energy non-conserving perturbations,
a planetary system could reach a self-organized periodic state from
arbitrary initial conditions. Their model, however, was simplistic
and could not properly simulate our solar system. Moreover, no simulation
using more than 6 identical planets was stable.

Our proposed empirical planetary model of the solar system is rather
peculiar because it suggests a planetary self-organization mechanism
that does not directly involve the traditional planetary commensurability
models based on whole number ratios of orbital periods, as usually
proposed in the literature \citep[cf.: ][]{Aschwanden,Peale}. In
fact, the $a^{2/3}$ metric cannot be directly interpreted using the
third law of Kepler, Eq. \ref{eq1}, because the latter links the
orbital periods to a $a^{3/2}$ metric. Moreover, as demonstrated
in section 3, the orbital-period metric does not yield ratios between
contiguous planets of our solar system that could be expressed by
consonances. Thus, our solar system is not gravitationally self-organized
like, for example, the Trappist-1 solar system.

Eqs. \ref{eq11} may suggest an alternative orbital self-organization
process that could involve gravity accelerations, and space and volume
ratios instead of the orbital period ones. For example, by assuming
that the orbits are circular (so that the semi-major axis coincides
with the orbital radius, $a_{1}=R_{1}$ and $a_{2}=R_{2}$), Eq. \ref{eq5}
with $k=2/3$ can be rewritten as: 
\begin{equation}
\left(\frac{a_{1}}{a_{2}}\right)^{\frac{2}{3}}=\left(\frac{R_{1}}{R_{2}}\right)^{\frac{2}{3}}=\left(\frac{m_{1}}{m_{2}}\frac{F_{2}}{F_{1}}\right)^{\frac{1}{3}}\approx f\label{eq18}
\end{equation}
where $m_{1}$ and $m_{2}$ are the masses of the two adjacent planets,
and $F_{1}$ and $F_{2}$ are the gravitational forces that attract
them toward the Sun ($F=GM_{\odot}m/R^{2}$); for each couple of adjacent
planets, the frequencies $f$ assume one of the values $2^{n/12}$
with $n$ = 4, 5, 7 and 8, or, alternatively, $f$ = 5/4, 4/3, 3/2
and 8/5. Thus, our result and Eq. \ref{eq18} indicate that the cube
root of the ratio between the centripetal orbital acceleration of
adjacent planets of the solar system can be interpreted as musical
tones and, more specifically, as consonances.

A 2/3rd power of an orbital radius could also be interpreted as a
geometrical transformation of an ellipsoid of radius $R_{e}$ and
fixed height $H$ into an equal volume sphere of radius $R_{s}$ according
to the equation $R_{s}=\sqrt[3]{HR^{2}}$. In fact, the solar system
is made of a planetary disk and its geometry could be approximated
by an ellipsoid with a given height $H$. Thus, each planetary orbit
could be associated with an ellipsoid with orbital radius $R$ and
a constant height $H$, and could be transformed into a sphere with
radius $R'=\sqrt[3]{hR^{2}}$. Our results would then imply that the
ratios of spherically transformed orbital radii of adjacent planets
of the solar system express musical consonances as: 
\begin{equation}
\left(\frac{R_{1}}{R_{2}}\right)^{\frac{2}{3}}=\frac{R'_{1}}{R'_{2}}\approx f.\label{eq19}
\end{equation}
Table \ref{tab:1} reports the $R'$ values for each planet.

We also observe that in physics, equations where cube frequencies
appear are not frequent, but one of them is the Planck's law (or its
Wien's approximation) describing the spectral density of electromagnetic
radiation emitted by a black body in thermal equilibrium at a given
temperature T \citep{Planck}. Finally, it is interesting to note
that the operation needed to obtain the above result from planetary
distances -- raising them to the 2/3 power -- is somehow specular
to the operation which relates planetary distances to orbital periods
by Kepler's third law, raising them to the 3/2 power (Eq. \ref{eq1}).

On whether the above or alternative analogies might yield a physical
relation between the relatively stable orbits of the solar system
and the distribution of gravitational energy in it linked to a 2/3rd
power of the orbital radii of the planets, is left to future investigations.

\section{Conclusion}

An interesting feature of the solar system is its specular-reflection-like
architecture which is made of four inner terrestrial planets (Mercury,
Venus, Earth and Mars) and four outer gas-giant planets (Jupiter,
Saturn, Uranus and Neptune) divided by the asteroid belt. No other
exoplanetary system similar to our has been discovered yet.

We have shown the Geddes--King-Hele equations for mirror symmetries
among the distances of the planets, when raised to the 2/3rd power,
express values that are very close to the simple ratios found in the
harmonic consonances of the 12-TET and 12-TJI tuning systems used
in Classical and Western music. This result contradicts the brief
critique of \citet{Abhyankar} that there is ``nothing particularly
musical'' in such equations. Of course, herein, we intend for the
word ``musical'' to relate to the presence of the ratios found in
Classical tuning systems which have specific mathematical properties.

Geddes and King-Hele noted the mirror symmetries but not the scaling
that we highlighted in our equations. This result further contradicts
\citet{Abhyankar}'s claim that such equations cannot tell us anything
``about the origin of the solar system or its stability''. In fact,
it appears that our solar system could be interpreted by Eq. \ref{eq:25}
(depicted in Figure \ref{fig6}) or Eq. \ref{eq:26} that relates
the ratios of planet pairs mirrored by the asteroid belt as a series
weighted by increasing powers of 2 of the Pythagorean tone \emph{epogdoon}
(the 9/8 ratio).

The orbital radii of the inner planets can be predicted from those
of the outer ones, and vice versa, with a precision that is about
three times superior to that of the harmonic orbit resonance model
recently proposed by \citet{Aschwanden}. In fact, it shows just a
0.8\% average error (that is an accuracy larger than 99\%) against
a 2.5\% error of the alternative method. In addition, the probability
of finding only musical consonances among such adjacent ratios has
a p-value $<0.1$\%, which makes it improbable that this is a random
result. Furthermore, our model could be extended to predict the inner
gap structure of the asteroid belt as well as the transneptunian objects.

Furthermore, Eq. \ref{eq21} (or Eq. \ref{eq22}) show that the coefficient
18 in Eq. \ref{eq14} is directly linked to the 3:1 and 7:3 resonances
with Jupiter that, by virtue of its large mass, has likely played
a decisive role in the orbital architecture of the solar system. This
main role seems confirmed in Figure \ref{fig5}A where the planetary
predictions of Eq. \ref{eq11} based on Jupiter (blue curve with circles)
are well balanced among the other series. The two cited resonances
characterize the main Kirkwood gaps of the asteroid belt. Thus, although
the physics behind such a result is not determined yet, these empirical
relations do not appear to be coincidental.

We also determined that for exponents $k$ close to $2/3$ there is
a convergent minimum both in the average and maximum error between
our proposed planetary metric and both the 12 musical tones and the
7 harmonic consonances. More specifically, for the solar system, such
planetary ratios are represented by harmonic musical consonances that
assume frequency values equal to $2^{n/12}$ with $n$ = 4, 5, 7 and
8, or, alternatively, 5/4 (Major Third), 4/3 (Perfect Fourth), 3/2
(Perfect Fifth) and 8/5 (Minor Sixth). Interestingly, the seven planets
of the Trappist-1 solar system (labeled b, c, d, e, f, g and h) present
a set of approximate orbital resonance ratios in the periods of adjacent
planets (from b$\leftrightarrow$c to g$\leftrightarrow$h) that includes
the same consonances: these are 8:5, 5:3, 3:2, 3:2, 4:3, 3:2 \citep[cf.][]{Agol,Gillon,Tamayo},
which correspond to the tones $Ab$, $A$, $G$, $G$, $F$ and $G$
(with C as a reference tone). Thus, we suggest that quasi-stable orbital
systems could be characterized by standard whole number ratios as
those that characterize the musical consonances. However, these ratios
can involve physical observables other than the orbital periods. Therefore,
alternative and/or complementary orbital metrics should be considered
for describing orbital systems.

In fact, mean motion resonances, in which the orbital periods or mean
angular velocities of planetary bodies are in ratios of small integers,
are commonplace in planetary systems, both in our own solar system
and in exoplanetary systems. The Trappist system that we mention is
a good example, while in our own solar system, a whole network of
mean motion resonances exist among the inner satellites of Saturn,
for example, with many other examples existing elsewhere \citep[e.g.][]{Aschwanden}.
These relationships are today well-understood as they satisfy Kepler's
third law and can be easily explained within the context of the laws
of planetary motion based on Newtonian gravity. The physical mechanisms
underpinning them, together with the secular and tidal evolution processes
which bring them about are well-established, and astrophysicists have
a good understanding of the interplay between regular and chaotic
motion which is fundamental to these (and actually to some degree
all) dynamical systems. However, such findings do not exclude the
possibility of alternative physical forms of self-organization of
orbital systems which are today still unknown or have not yet been
investigated.

For our solar system, the consonant ratios among adjacent planets
emerge when the ellipsoid orbital radii are transformed into equal-volume
spherical radii using the exponent $k=2/3$, but for the Trappist-1
system, the orbital radii are to be transformed into periods using
the exponent $k=3/2$. Thus, it appears that what happens for the
solar system cannot be easily explained in terms of the usual Newtonian
motion resonance approaches. The evidence suggests that the exponent
$k$ could differ for different orbital systems and the found $k=2/3$
exponent may express an alternative metric capable of producing a
self-organizing orbital structure.

These different kinds of harmonic structures could in the future be
properly understood and classified as more and more exoplanetary systems
are discovered. This task is made more difficult today because testing
for a relationship such as Eq. \ref{eq:25} in exoplanetary systems
may not be possible until our knowledge of them is complete. In fact,
it is difficult to fully characterize detailed orbital information
for all the large and small planets, in addition to possible asteroid
belts in distant exoplanetary systems. The challenge for future research
would be to justify the proposed metric on physical grounds or to
find a better physical explanation for the self-organization of the
solar system, which, however, is today a matter of debate.

In conclusion, the ratios of the orbital radii of adjacent planets
of our solar system, when raised to the 2/3rd power, express the simple
ratios found in harmonic musical consonances and can be expressed
by a simple, elegant, and highly precise equation that reveals scaling
and mirror-like symmetries of its planetary orbital distribution relative
to the asteroid belt, whose inner structure is also predicted by the
same model depicted in Figure \ref{fig6}. Eqs. \ref{eq:25} and \ref{eq:26}
suggest that the orbital scaling-mirror symmetries of the solar system
could be expressed by the Pythagorean \emph{epogdoon} (the tone ratio
9/8) and its addition with up to six octaves. Furthermore, the ratio
9/8 is closely related to the 3:1 and 7:3 resonances of Jupiter that
shape the asteroid belt (Eq. \ref{eq22}), which indicates the primary
role played by Jupiter in organizing the planetary orbits of the solar
system.

\begin{figure*}[!t]
\includegraphics[width=1\textwidth]{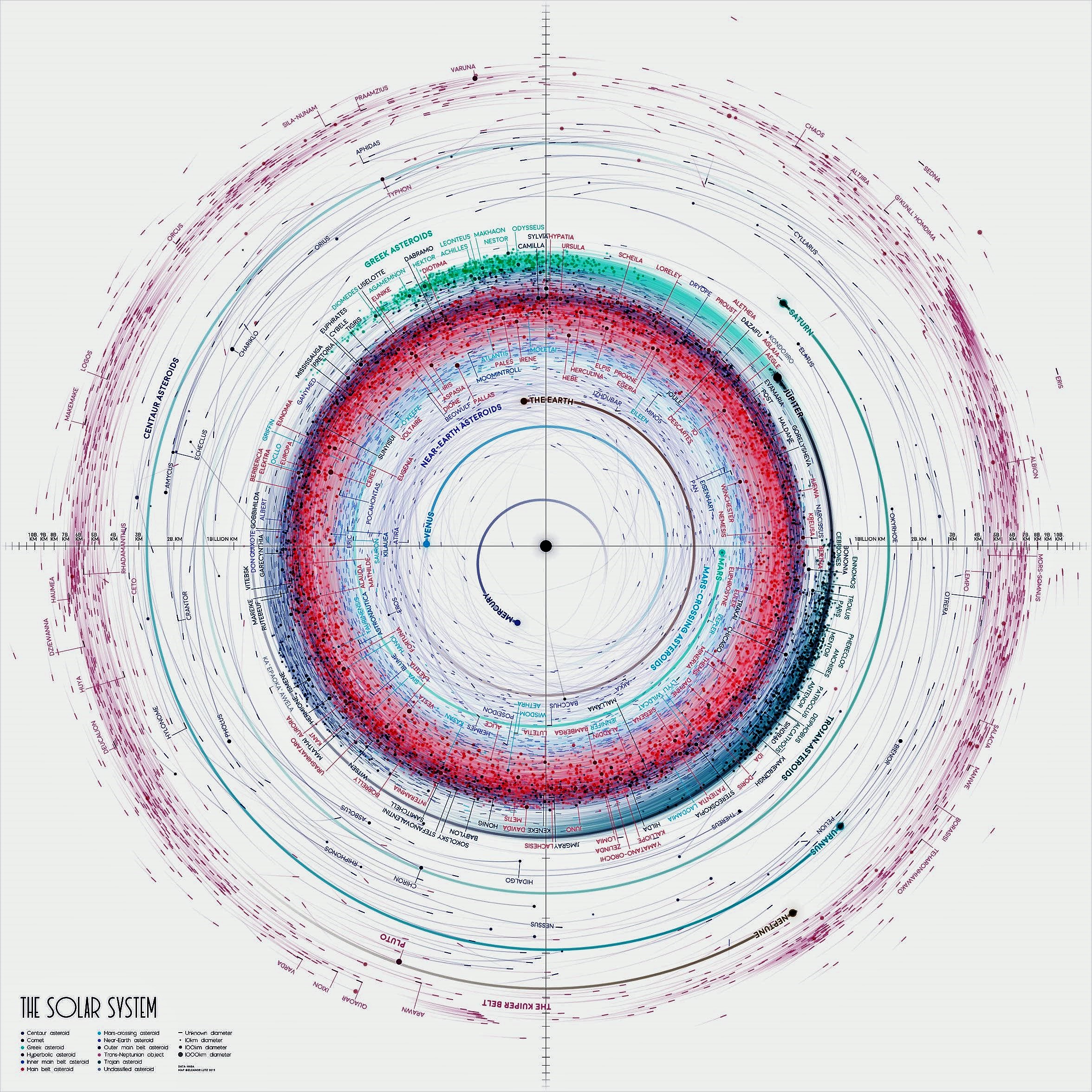}\caption{A nice presentation of the solar system. This map shows in a logarithmic
scale the orbit paths of the planets of the solar system together
with those of about 18,000 asteroids and comets larger than 10 km
in diameter. Adapted from \citet[CC BY-NC-ND 4.0]{Lutz}.}
\end{figure*}

The mathematical correlation which we presented between idealized
musical ratios and planetary data is very similar to what Johannes
Kepler was seeking when he published his \emph{Harmonices Mundi} in
1619. Our result further indicates that the orbital movements of the
major bodies of the solar system are likely highly organized. In this
regard, we would also like to point out that aesthetic perception
of patterns in our surroundings is a fundamental dimension of the
human culture, and it has been crucial in the development of a scientific
understanding of the natural world. Thus, our empirical model could
lead to the future discovery of important dynamical structures of
orbital systems, which today are still unknown. The final paragraph
of Geddes \& King-Hele's original 1983 paper is worth quoting: 
\begin{quote}
``\emph{The significance of the many near-equalities is very difficult
to assess. The hard-boiled may dismiss them as mere playing with numbers;
but those with eyes to see and ears to hear may find traces of `something
far more deeply interfused' in the fact that the average interval
between the musical notes emerges as the only numerical constant required
- a result that would surely have pleased Kepler}.''
\end{quote}
\onecolumn

\section*{Appendix}

Eq. \ref{eq:26} can be rewritten also in the following forms that
highlight better how each pairs of planets is directly linked to the
Pythagorean \emph{epogdoon} ratio 9/8 (or 8/9):

\begin{equation}
\frac{1}{64}\left(\frac{a_{Er}}{a_{Sz}}\right)^{\frac{2}{3}}\approx\frac{1}{32}\left(\frac{a_{Pl}}{a_{Vu}}\right)^{\frac{2}{3}}\approx\frac{1}{16}\left(\frac{a_{Ne}}{a_{Me}}\right)^{\frac{2}{3}}\approx\frac{1}{8}\left(\frac{a_{Ur}}{a_{Ve}}\right)^{\frac{2}{3}}\approx\frac{1}{4}\left(\frac{a_{Sa}}{a_{Ea}}\right)^{\frac{2}{3}}\approx\frac{1}{2}\left(\frac{a_{Ju}}{a_{Ma}}\right)^{\frac{2}{3}}\approx1\left(\frac{a_{7:3}}{a_{3:1}}\right)^{\frac{2}{3}}\approx\frac{9}{8}
\end{equation}

\begin{equation}
64\left(\frac{a_{Sz}}{a_{Er}}\right)^{\frac{2}{3}}\approx32\left(\frac{a_{Vu}}{a_{Pl}}\right)^{\frac{2}{3}}\approx16\left(\frac{a_{Me}}{a_{Ne}}\right)^{\frac{2}{3}}\approx8\left(\frac{a_{Ve}}{a_{Ur}}\right)^{\frac{2}{3}}\approx4\left(\frac{a_{Ea}}{a_{Sa}}\right)^{\frac{2}{3}}\approx2\left(\frac{a_{Ma}}{a_{Ju}}\right)^{\frac{2}{3}}\approx1\left(\frac{a_{3:1}}{a_{7:3}}\right)^{\frac{2}{3}}\approx\frac{8}{9}
\end{equation}

\subsection*{Data availability}

The calculations presented herein are based on the orbital data listed
in Table \ref{tab:2} which are provided by NASA: \href{https://nssdc.gsfc.nasa.gov/planetary/factsheet/}{https://nssdc.gsfc.nasa.gov/planetary/factsheet/}

\subsection*{Contributions}

MB contributed to the interpretation of the results based on music
theory; NS developed the scientific and astronomical interpretation
of the results, wrote and organized the paper. Both authors share
first authorship, contributed to the discussion and have read and
edited the text of the final manuscript.

\subsection*{Competing interests}

The authors declare that the research was conducted in the absence
of any commercial or financial relationships that could be construed
as a potential conflict of interest.

\end{document}